\documentclass[fleqn,usenatbib]{mnras}

\usepackage{newtxtext,newtxmath}
\usepackage{hyperref}
\hypersetup{colorlinks=true,linkcolor=blue,citecolor=blue,filecolor=blue,urlcolor=blue,pdfpagelabels=false}

\usepackage[T1]{fontenc}

\DeclareRobustCommand{\VAN}[3]{#2}
\let\VANthebibliography\thebibliography
\def\thebibliography{\DeclareRobustCommand{\VAN}[3]{##3}\VANthebibliography}
\def\dof{{\tt DOF}}
\def\uwe{{\tt UWE}}
\def\w{$\omega$}
\def\noise{$\epsilon$}
\def\chandra{{\em Chandra}}
\def\gaia{{\em Gaia}}
\def\erosita{{\em eROSITA}}
\def\msun{$M_\odot$}
\def\ha{H$\alpha$}
\def\mbh{$M_{\rm BH}$}

\def\p{$\pm$}
\def\nh{$N_{\rm H}$}
\def\porb{$P_{\rm orb}$}
\def\bprp{$B_{\rm p}$\,--\,$R_{\rm p}$}
\def\gabs{$G_{\rm abs}$}
\def\ltsim{\mathrel{\hbox{\rlap{\hbox{\lower4pt\hbox{$\sim$}}}\hbox{$<$}}}}
\def\gtsim{\mathrel{\hbox{\rlap{\hbox{\lower4pt\hbox{$\sim$}}}\hbox{$>$}}}}

\def\pertenmille{\ensuremath{{}^\text{o}\mkern-5mu/\mkern-3mu_\text{ooo}}}

\usepackage{graphicx}	
\usepackage{amsmath}	
\usepackage{amssymb}	

\title[X-ray sources with astrometric excess noise]{Astrometric excess noise in Gaia DR2 and the search for X-ray emitting binaries}
\author[P. Gandhi et al.]{P.\,Gandhi,$^{1,\,2}$\thanks{E-mail: poshak.gandhi@soton.ac.uk (PG)} 
D.A.H.\,Buckley,$^{3}$
P.A.\,Charles,$^1$
S.\,Hodgkin,$^{4}$
S.\,Scaringi,$^{5}$
C.\,Knigge,$^1$ 
A.\,Rao$^1$
\\
$^{1}$School of Physics \& Astronomy, University of Southampton, Highfield SO17 1BJ\\
$^{2}$Inter-University Centre for Astronomy \& Astrophysics, Post Bag 4, Ganeshkhind, Pune, 411007, India\\
$^{3}$South African Astronomical Observatory, PO Box 9, Observatory 7935, Cape Town, South Africa\\
$^{4}$Institute of Astronomy, Madingley Road, University of Cambridge, Cambridge, CB3 0HA\\
$^{5}$Department of Physics, Durham University, South Road, Durham, DH1 3LE\\
}

\date{Submitted 2020 Aug 27}

\pubyear{2021}

\begin{document}
\label{firstpage}
\pagerange{\pageref{firstpage}--\pageref{lastpage}}
\maketitle

\begin{abstract}
Astrometric noise (\noise) in excess of parallax and proper motion could be a signature of orbital wobble (\w) of individual components in binary star systems. The combination of X-ray selection with astrometric noise can then be a powerful tool for identifying accreting binaries. Here, we mine the \gaia\ DR2 catalogue for Galactic sources with significant values of astrometric noise over the parameter space expected for known and candidate X-ray binaries (XRBs). Cross-matching our sample with the \chandra\ Source Catalogue CSC\,2.0 returns a primary sample of $\approx$\,1,500 X-ray sources with significant \noise, constituting $\approx$\,0.04\% of the initial \gaia\ sample. By contrast, the fraction of matched X-ray sources in a control sample with smaller \noise\ is a factor of about 7 lower. The primary sample branches off the main sequence much more than control objects in colour-mag space, shows a distinct spatial distribution in terms of Galactic latitudes, comprises more objects with an H$\alpha$ excess and larger X-ray-to-optical flux ratios, and includes a higher fraction of known binaries, variables and young stellar object class types. However, for individual XRBs with known system parameters, \noise\ can exceed expectations, especially for small $\omega$. It is likely that other factors (possibly attitude and modelling uncertainties, as well as source variability) currently dominate the observed noise in such systems. Confirmation must therefore await future \gaia{} releases. The full X-ray matched catalogue is released here to enable legacy follow-up. 
\end{abstract}




\section{Introduction}

There are only about 25 known stellar-mass black holes in the Milky Way with confirmed mass estimates. All of them lie in binary systems where spectroscopic radial velocity variations of the companion star have been used to confirm the presence of massive compact objects \citep{blackcat, watchdog}. By contrast, the Galaxy is expected to host anywhere between $\sim$\,10$^3$--10$^8$ stellar-mass black holes in binary systems \citep[e.g. ][]{pfahl03,tetarenko16}. This population should be dominated by non-accreting systems and incipient black hole X-ray binaries (BHXBs) with only a handful of recent well studied systems  \citep{thompson19, tetarenko16, rivinius20}. An interesting recent highlight in this field was LB-1, with a proposed \mbh\,=\,70\,\msun\ \citep{liu19_lb1}. Though now believed to be much lighter \citep[e.g. ][]{eldridge20}, its discovery accelerated efforts to understand the space density of massive quiescent BHXBs. Another example is the report of a putative black hole in the triple system HR\,6819 \citep{rivinius20}, though this also remains controversial \citep{bodensteiner20, elbadry20}.

A relatively unexplored avenue of research is the ability of \gaia\ to detect {\em new} candidate BHs in binaries. This is made possible by \gaia's exquisite astrometric precision. In particular, orbital motion of the companion star in a binary system will result in `astrometric orbital wobble (\w)', over and above the parallax and proper motion locus for any given system. This can manifest as an `astrometric excess noise (\noise)', one of the parameters reported in the recent data release 2 (DR2; \citealt{gaiadr2}). \noise\ is defined in DR2 as the excess uncertainty that must be added in quadrature to obtain a statistically acceptable astrometric solution \citep{gaiadr2, lindegren12}.\footnote{\url{https://gea.esac.esa.int/archive/documentation/GDR2/Gaia_archive/chap_datamodel/sec_dm_main_tables/ssec_dm_gaia_source.html}}. In the early data releases, \noise\ includes instrument and attitude modelling errors that are statistically significant and could result in large values of \noise. Thus, a detailed investigation of \noise-based selection is warranted, and this is what we carry out herein. 

Several recent theoretical works have highlighted the feasibility of large surveys, including \gaia\ astrometry \citep{breivik17} and microlensing with missions such as TESS \citep{masuda19}, to uncover large new populations of black holes in binary orbits. Massive spectroscopic surveys are also beginning to probe this territory through brute force blind searches for radial velocity variations characteristic of massive compact objects \citep{pricewhelan20}. Furthermore, large multiwavelength surveys such as \erosita\ in X-rays \citep{erosita} and ngVLA in the radio \citep{maccaronengvla} will be instrumental in confirming the nature of newly identified candidates. Thus, there are enormous synergies waiting to be explored in this field. 

We begin by exploring the parameter range of interest in terms of XRB candidates and expected astrometric wobble in \S\,\ref{sec:understanding}, and then define a data mining strategy to identify  promising candidates (\S\,\ref{sec:data}). Though our initial aim was to concentrate solely on searches for new BHXB candidates, the diversity of systems that we uncover in the Results section (\S\,\ref{sec:results}) herein enabled us to broaden the study to a wider range of object types, and shows that \noise\ selection could be a powerful tool of general utility. The Discussion section (\S\,\ref{sec:discussion}) delves into the consequences and caveats of our findings, before the work is summarised. There is also an online Appendix including the full X-ray cross-matched catalogue of objects, extra details on object classification, in addition to a few tests and simulations described in the main text. 

Unless otherwise stated statistical uncertainties quoted throughout the paper refer to 68\%\ confidence Poisson or binomial limits, and are appropriate for small number statistics \citep{gehrels86}.

\section{Astrometric wobble and excess noise parameter space}
\label{sec:understanding}

\begin{figure*}
\hspace*{-1cm}
	\includegraphics[angle=90,width=\columnwidth]{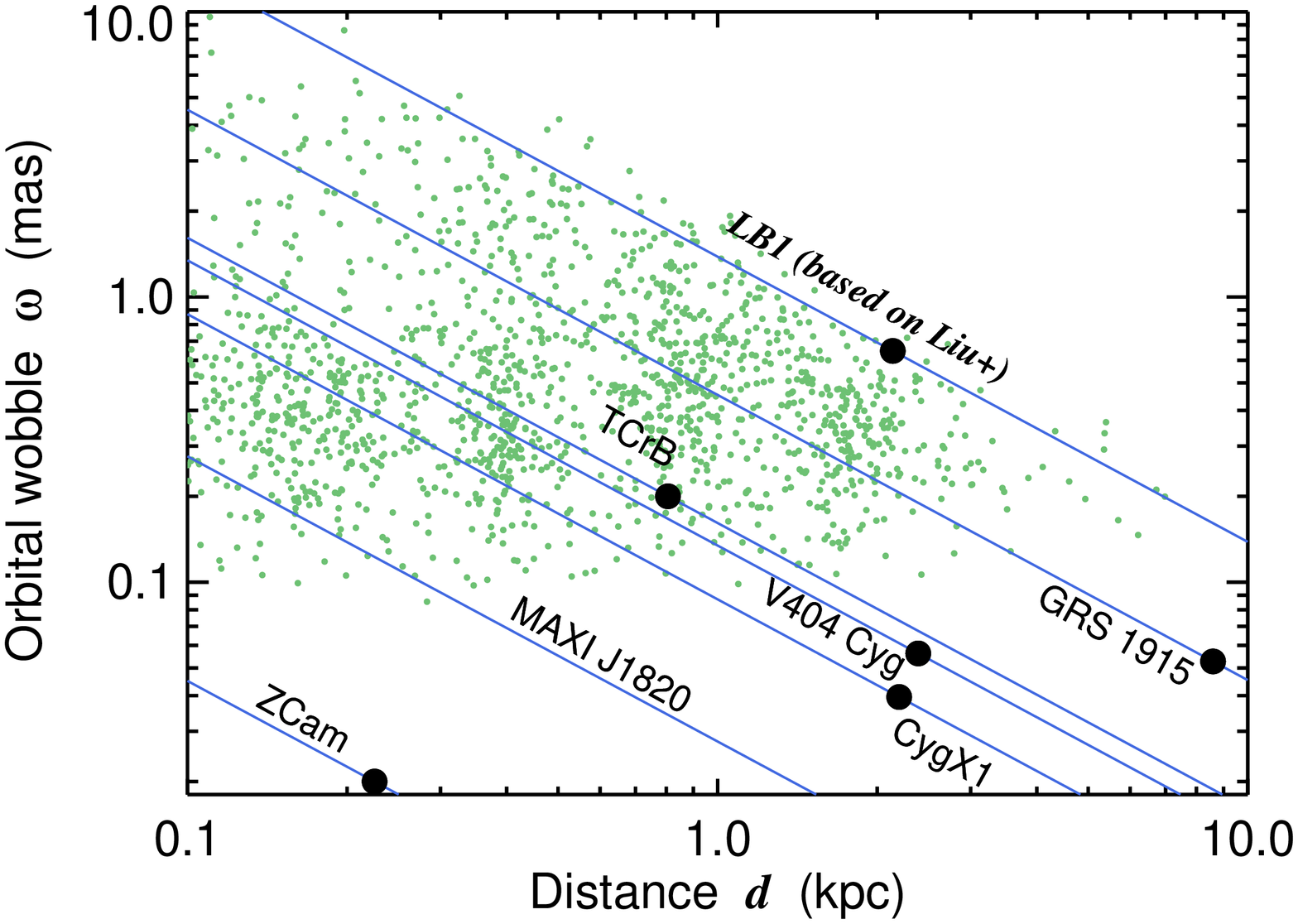}
    \includegraphics[angle=90,width=\columnwidth]{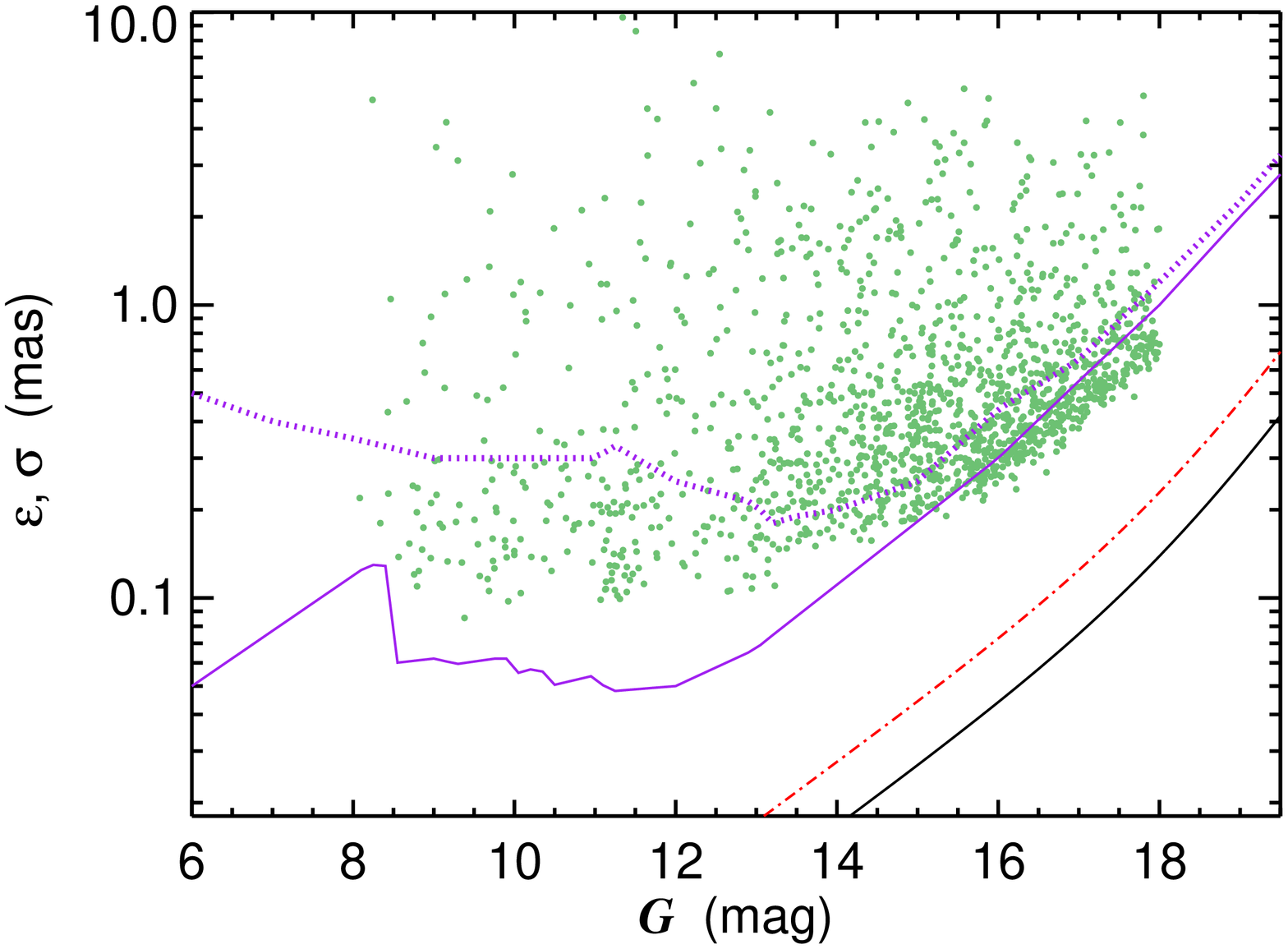}
\caption{{\bf (Left)} Expected maximal orbital wobble (\w) as a function of source distance $d$ for a few well-known candidate and known interacting binaries. The black dots denote the expected wobble at the actual source distance. LB1 is likely to be much less massive than initially indicated (see text), but serves as a useful comparison. {\bf (Right)} The astrometric noise floor is magnitude dependent and is denoted in this panel in three ways: a robust estimate of the epochwise standard deviation is shown by the purple dotted curve, and the formal uncertainties are approximated by the continuous purple curve \citep{lindegren18}. The cumulative end-of-mission (5 years) and DR2 uncertainties are denoted by the black continuous line and the dashed red line, respectively. In both panels, green dots show the DR2 {\tt excess\_noise} parameter \noise\ (as a proxy for \w) for our primary sample of $\sim$\,1,500 X-ray sources (see \S\,\ref{sec:chandra}).}
\label{fig:wobble}
\end{figure*}

Astrometric wobble (\w) is defined here simply as the maximal projected areal half angle swept by the companion star over its orbit, if the observed flux is dominated by the companion star (this should be mostly true in quiescent, non-accreting, and high-mass XRBs). 

\begin{equation}
    \omega = \frac{a_2}{d}
\end{equation}

\noindent
where $a_2$ is the semi-major axis of the companion and $d$ the source distance. In terms of other observed and system orbital parameters, this may be expressed as

  \begin{equation}
    \label{eq:wobble}
\omega=\frac{G M_{1}}{K_2^2 \left( 1+ q \right)^2d} \sin^2 i 
\end{equation}

\noindent
with $K_2$ the companion velocity amplitude which will be dependent on the binary inclination $i$, and $q$ the mass ratio of the companion to the primary $\frac{M_2}{M_1}$. For simplicity, a circular orbit is assumed here,  since interacting binaries (our core targets of interest) are expected to circularise rapidly.  While this is not the case for the (often highly) eccentric Be X-ray systems (see e.g. \citealt{Reig11}), virtually all known Be X-ray sources have NS compact objects, so the astrometric wobble of the (much) more massive Be donor is expected to be small. Fig.\,\ref{fig:wobble} denotes \w\ as a function of distance for a few well known XRBs, in which \w\ is expected to lie in the range of $\sim$\,0.01--1.0\,mas. Details of most of these can be found in \citet{gandhi19},\footnote{the only difference being that $\omega$ was defined in \citet{gandhi19} to be the full swept parallax angle over an orbit, twice the value defined here.} with a few others based up system parameters from \citet[][Z\,Cam]{robinson76}, \citet[][T\,CrB]{fekel00} and \citet[][LB1]{liu19_lb1}. For the case of T\,CrB, we were additionally able to account for the expected shift in the centre-of-light due to the known flux contributions of the primary and the secondary components of the binary (both of which are luminous); details may be found in the Appendix.  Furthermore, as already mentioned, LB1 and similar recent discoveries remain controversial and are likely to be lighter than initially envisaged. But the presented trends nevertheless serve as guides to the potential discovery of other similar systems.

Astrometric noise \noise\ represents an additional intrinsic scatter term in the DR2 pipeline astrometric solution. This is the value that needs to be added in quadrature to the formal statistical uncertainties ($\sigma$) in order to make the solution statistically acceptable, effectively making the reduced sum-of-squared-weighted-residuals equal unity \citep{lindegren12}. 

Since $\omega$ represents a deviation from the nominal parallax locus, it is equivalent to an excess scatter term, and thus conceptually similar to \noise, if no other noise term contributes. In such a case, the expectation value of \noise\ should approximate $\hat{\omega}$\,=\,$\omega$/$\sqrt{2}$ in the limit of perfect sampling. However, in early \gaia\ releases, \noise\ absorbs instrumental as well as attitude modelling errors that are likely to be statistically significant. The excess noise terms are globally adjusted to match the weighted sum of residuals to the number of degrees of freedom \citep{lindegren18}, so there could be some potential degeneracy between the magnitudes of the noise terms, and they need not scale directly and strictly with $\omega$ for individual objects. Thus, one must study the distributions of these parameters in a controlled manner before extracting scientifically useful inferences \citep{lindegren12, luri18, gaiadr2}, and we will do so in later sections. 

Nevertheless, first insight into expectations may be obtained by examining order-of-magnitude estimates of the mission uncertainties. The per-epoch astrometric uncertainties remain unavailable in the current data releases. But it is possible to use estimates published by the \gaia\ team of the formal astrometric uncertainties as a first comparison. These brightness dependent estimates from \citet{lindegren18} are shown in Fig.\,\ref{fig:wobble}b, illustrating both the formal and robust scatter estimates of single epoch observations, as well as long-term estimates of the DR2 and end-of-mission uncertainties. If these estimates remain valid, then useful constraints may be expected well below a level of $\sim$\,1\,mas across the bulk of the astrometric sensitivity regime of \gaia, but reaching below the $\approx$\,0.1\,mas level will be restricted to brighter sources. 

For the reader interested in the details of how \gaia\ sampling and its effect on \noise\ measurement, the Appendix presents a simulation of DR2 observations of one XRB, T\,CrB, with discussion on caveats therein.

\section{Data}
\label{sec:data}

\subsection{Mining \gaia}

Based upon the orbital wobble expected for XRBs, we used the following criteria for investigation, taking \noise\,$\ge$\,0.01\,mas as a starting threshold for this sample, together with significance of 5 for the excess noise. The median mag of BHXBs with five-parameter astrometric solutions is $G$\,$\approx$\,17.4, and one of the brightest confirmed BHXB (Cyg\,X--1) has $G$\,=\,8.5 \citep{gandhi19}. We thus chose a mag range of  8\,$\le$\,$G$\,$\le$\,18. This  avoids ballooning the sample at the faint end where statistical errors are expected to dominate, and also circumvents issues with systematic uncertainties for very bright sources.\footnote{\url{https://www.cosmos.esa.int/web/gaia/dr2-known-issues\#PhotometrySystematicEffectsAndResponseCurves}} A distance range of 0.1--10\,kpc was examined, requiring a significant parallax (distance) measurement in order to try and assess the nature of the source population, as discussed in the following section. A minimum threshold on the number of visibility periods is used to ensure adequate sampling in time in the astrometric fit. The DR2 ADQL \citep{adql} query for our primary sample is:
~\newline

\noindent
{\tt
SELECT *\\
FROM gaiadr2.gaia\_source\\
WHERE (parallax\_over\_error $>$ 4) AND \hspace*{0.25cm}(visibility\_periods\_used $>$ 10) AND\\
\hspace*{0.25cm}(parallax $<$ 10.) AND (parallax $>$ 0.1) AND\\ 
\hspace*{0.25cm}(phot\_g\_mean\_mag $>$ 8) AND (phot\_g\_mean\_mag $<$ 18)\\ 
\hspace*{0.25cm}AND (astrometric\_excess\_noise > 0.01)\\
\hspace*{0.25cm}AND (astrometric\_excess\_noise\_sig > 5).
}
~\newline

\noindent
We next defined a control sample for cross-comparison. The DR2 ADQL query for this control sample is identical to the above except for the excess noise selection criterion, because this is our main parameter of interest. We thus use a complementary criterion, as follows:

{\tt
(astrometric\_excess\_noise < 0.01).
}
~\newline

\noindent
The above selection criteria are termed `G1', for the respective \gaia\ primary and control samples. 

Good astrometric fits require sources to be free from confusion and blending with close neighbours. We thus next excluded all objects with any detected DR2 near-neighbours (criterion G2). A radius of 5\,arcsec was adopted for our near-neighbour threshold. This is fairly conservative, given that the nominal \gaia\ point spread function is concentrated well within 1\,arcsec.\footnote{\url{https://gea.esac.esa.int/archive/documentation/GDR2/Data_processing/chap_astpre/sec_cu3pre_cali/ssec_cu3pre_cali_psflsf.html}} But it ensures that we can focus on clean astrometric solutions for isolated objects in crowded regions of the Galactic plane. This exclusion was applied to both the the primary and the control samples.

Finally, we included the photometric quality criterion proposed by \citet[][see also \citealt{lindegren18, luri18, babusiaux18}]{evans18}:

\begin{equation}
    1.0 + 0.015(G_{\rm BP} - G_{\rm RP})^2 < E < 1.3 + 0.06(G_{\rm BP} - G_{\rm RP})^2
\end{equation}

\noindent
where $G_{\rm BP}$ and $G_{\rm RP}$ represent source mags in the $B_P$ and $R_P$ bands, respectively, and $E$ is the photometric excess flux factor. This removes sources with potential suspect photometry in these bands.  This (criterion G3) should allow robust examination of source types based on colour. 

We did {\em not} remove sources based upon the astrometric goodness of fit $\chi^2$, because our selections are defined around the need for sources to show deviations from standard single star fits.

\subsection{Cross-match with X-rays}

X-ray activity is a key signature of the presence of an interacting accreting binary, so we searched for archival X-ray detections of our collated samples. Quiescent BHXBs are expected to display low level accretion activity, with typical luminosities of up to $L_{\rm X}$\,$\sim$\,10$^{32}$ erg\,s$^{-1}$. NSXBs in quiescence can be even more luminous, on average \citep[e.g. ][]{reynoldsmiller11}. However, detection of X-rays by itself is not unambiguous proof of the presence of such an interacting binary, with other possibilities including magnetically active stars \citep[e.g. ][]{Gudel04}, colliding winds \citep[e.g. ][]{PIttard18}, and activity in young stellar objects \citep[e.g. ][]{feigelson99}. So care will be necessary when making final inferences.

The archival database that we used was the \chandra{} Source Catalogue (CSC; \citealt{csc}). This is one of the largest, and most sensitive databases in terms of broadband X-ray sky coverage, and delivers exquisite spatial resolution ($\ltsim$\,1\,arcsec on axis). High precision centroiding is critically important in crowded regions such as the Galactic plane, along which many of our sources will fall. The latest data release, CSC2.0 \citep{csc2},  covers approximately 550\,deg$^2$ (1.3\,\%) of the sky down to a point source sensitivity limit of 5\,counts. Assuming an accreting source characterised by a power-law spectrum with slope $\Gamma$\,=\,2,\footnote{Photon rate density $N(E)$\,$\propto$\,$E^{-\Gamma}$ at energy $E$.} this corresponds to a 0.5--7\,keV flux $F_{\rm X}$ of 6\,$\times$\,10$^{-15}$\,erg\,cm$^{-2}$\,s$^{-1}$ for the median CSC2 exposure time of 12\,ks.\footnote{\url{https://cxc.harvard.edu/csc/char.html}} This is an approximation based upon the latest response function\footnote{\url{https://heasarc.gsfc.nasa.gov/cgi-bin/Tools/w3pimms/w3pimms.pl}} and assuming a line-of-sight column density \nh\,=\,5\,$\times$\,10$^{21}$\,cm$^{-2}$, not atypical out to distances of a few kpc in the Galactic plane. The \chandra\ soft energy response has been degrading with time so it is likely that older observations were more sensitive, on average. Taking the above flux limit as a baseline for comparison, CSC2 should be able to detect XRBs out to a distance $d$ with luminosity greater than 

\begin{equation}
    L_{\rm X-ray}>7\times 10^{29} \left(\frac{d}{1\,\rm{kpc}}\right)^{2}\,{\rm erg\,s^{-1}\,\,\, [0.5-7\, keV]}.
\end{equation}

\subsection{\ha\ observations}

In order to obtain further robust indicators of the presence of accreting sources, we searched for detections of the \ha\ emission line. This is a classic optical signature of accretion in quiescent, viscously heated discs (e.g. \citealt{casares15}). We achieved this in two ways: (1) using a population approach, and also (2) a  dedicated follow-up of a handful of individual candidates. 

For the first (population approach), we cross-matched our X-ray detected samples against the large IPHAS catalogue \citep{iphas, scaringi18}. IPHAS (The INT Photometric \ha{} Survey of the Northern Galactic Plane) is a massive survey imaging the Northern Milky Way in visible light (\ha, $r$ and $i$) down to mag\,$>$\,20, using the Isaac Newton Telescope (INT) in La Palma. Based upon the typical seeing and crowding, we used a 2$\arcsec$ radius for cross-matching. Our results are described in Section\,\ref{sec:discussion} and the full updated results from IPHAS will be published in Fratta et al. (in prep.)

For the second (dedicated pointed  follow-up), we  observed three sources with the Southern African Large Telescope \citep[SALT;][]{Buckley2006}, using the fibre-fed High Resolution Spectrograph (HRS;  \citealt{Bramall2012,salthrs}). The observations were carried out on three nights, 2020 May 11 at 23:55 UT, 2020 Jun 01 at 19:12 UT, and 2020 Jun 02 at 19:15 UT\footnote{The times refer to candidates \#4, 5 and 3, respectively, from Table\,\ref{tab:lum}, discussed in Section\,\ref{sec:discussion}.} in thin cirrus conditions. Exposure times were, respectively 1000, 300 and 1000 sec. 
Spectra were obtained in the Low Resolution (LR; R\,$\sim$\,15,000) mode of HRS, with the red and the blue arms covering the wavelength regions 3800$-$5550\,\AA\ and 5450$-$9000\,\AA, respectively. The spectra were reduced using the weekly set of HRS calibrations, including ThAr arc spectra and QTH lamp flat-fields. 
Primary data reduction was done using the SALT science pipeline, {\tt PySALT} \citep{Crawford2016}, correcting for overscan, bias and gain correction. The spectral reductions were then undertaken using a MIDAS-based {\'e}chelle reduction package \citep[see details in][]{Kniazev2016}. The results are discussed in Section\,\ref{sec:discussion}. 
 
\section{Results}
\label{sec:results}

\begin{table}
    \centering
            \caption{Sample selection statistics
\label{tab:stats}}
    \begin{tabular}{lcr}
\hline
\hline
Criterion & Primary & Control\\
\hline
G1 & 8,108,098 & 80,892,284 \\
G1+G2 & 3,537,090 & 49,521,792\\ 
G1+G2+G3 & 3,430,862 & 49,466,387\\
$''$ (distance-matched) & \multicolumn{2}{c}{2,885,782} \\
&&\\
G1+G2+G3+X-ray & 1,689 & 3,750 \\
G1+G2+G3+$F_{\rm X}$\,$>$\,0 & 1,490 & 3,222 \\
$''$ (distance-matched) & 1,336 & 1,354 \\
&&\\
G1+G2+G3+X-ray+\ha & 171 & 292 \\ 
G1+G2+G3+X-ray+\ha-excess & 31 & 20\\ 
\hline
    \end{tabular}
\end{table}

\subsection{{\gaia\ sample statistics}}
\label{sec:fullsample}

Table\,\ref{tab:stats} lists the number of objects detected according to various selection criteria. Our \gaia/DR2 mining (criteria G1+G2+G3) resulted in over 3.4 million sources selected in the `primary' sample. These are sources with significant astrometric excess noise, clean photometry and no close neighbours. The corresponding `control' sample was much larger as expected, with just under 50 million sources. The exclusion of close neighbours has a substantial impact on the selected sample, cutting the initial selection (G1) of the samples by 40--60\,\%. Thereafter excluding the photometric quality criterion (G2) has a relatively modest effect of a few per cent on the sample size. 

\begin{figure*}
    \includegraphics[angle=0,width=1\columnwidth]{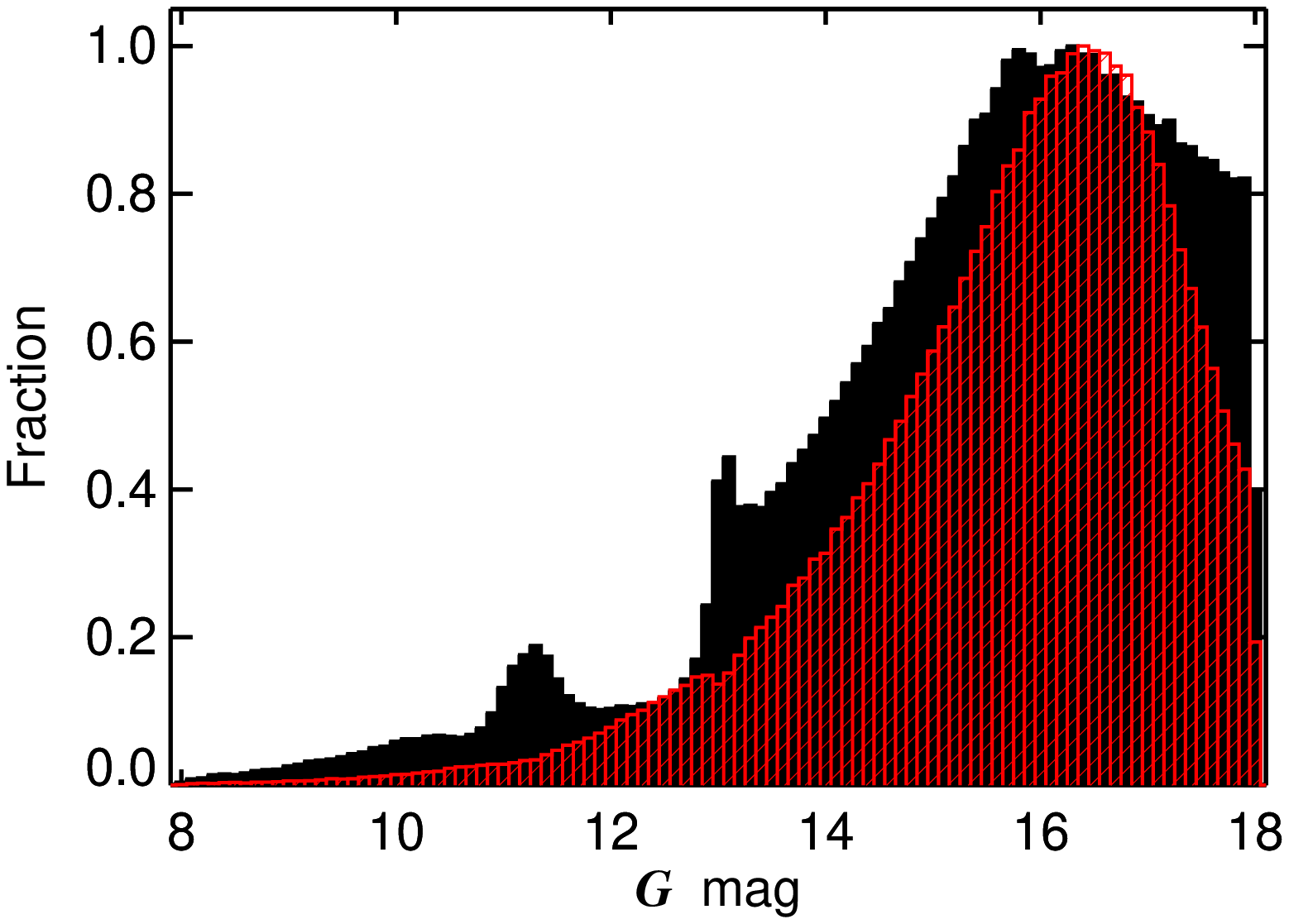}
    \hfill
    \includegraphics[angle=0,width=0.95\columnwidth]{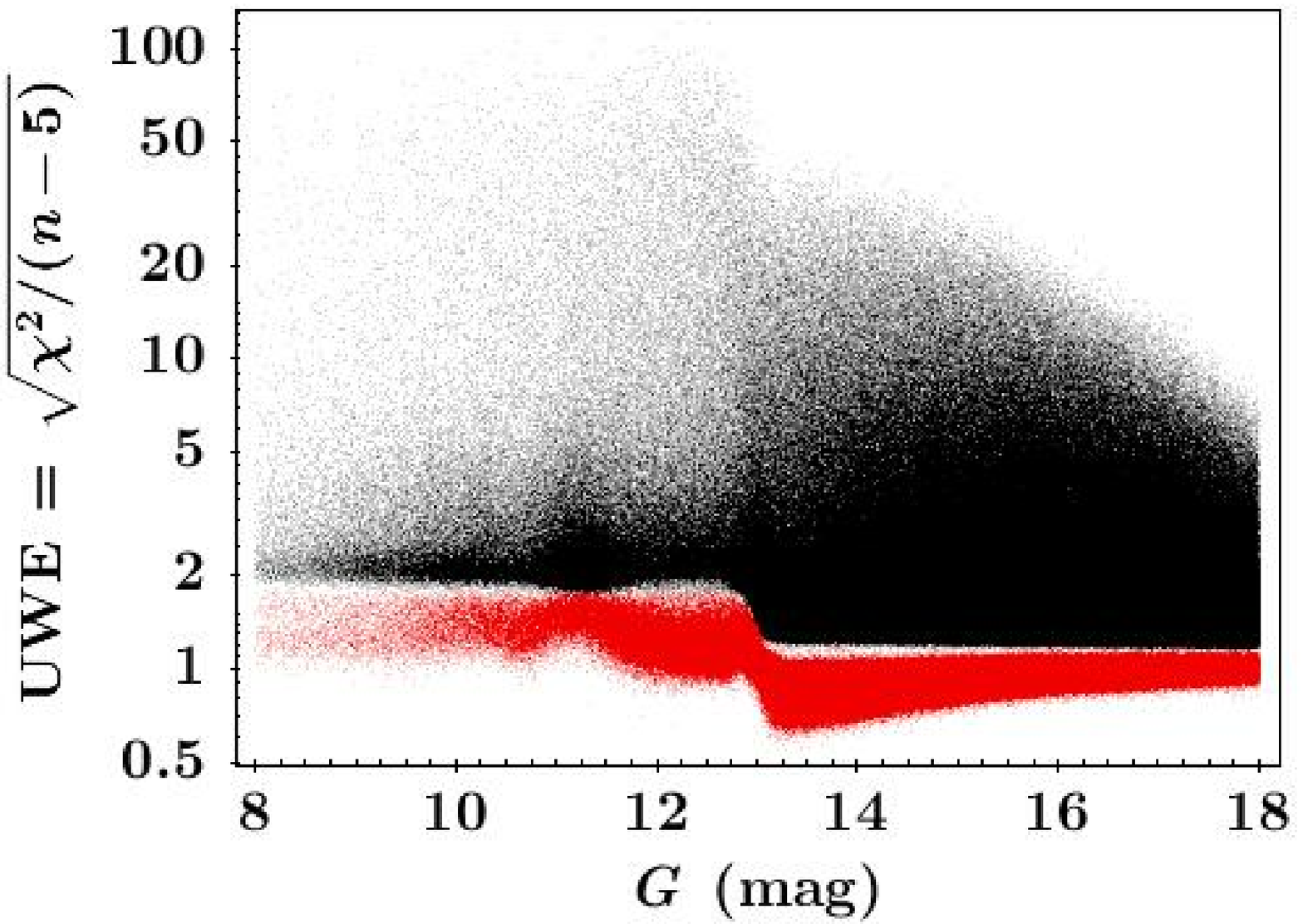} %
    \includegraphics[angle=0,width=1\columnwidth]{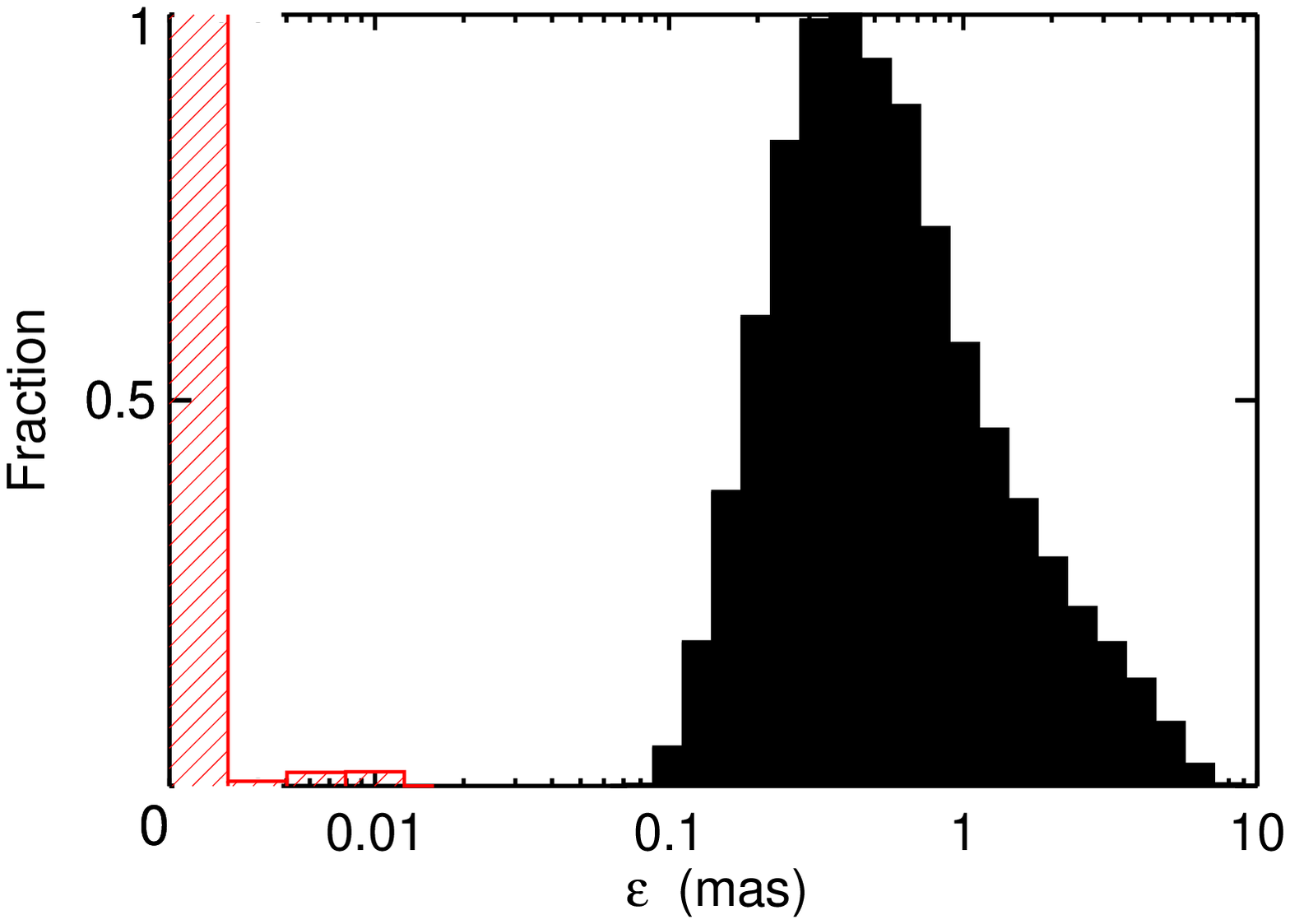}
    \caption{{\bf (a)} $G$ mag distributions of the primary (black) and control (red) samples. For both samples, a random selection of 1 million objects from the \gaia\ only selection (criteria G1+G2+G3) is used to construct the histograms. {\bf (b)} The unit weight error (UWE) distribution as a function of $G$ mag for the two samples. The primary sample shows substantially larger UWE values, as expected. In both panels, there are features around $G$\,$\approx$\,11 and 13\,mag, related to DR2 pipeline related artefacts (see text in Section\,\ref{sec:fullsample}). {\bf (c)} Histograms of {\tt astrometric\_excess\_noise} \noise. The vast majority of control sample objects show \noise\,=\,0, so their histogram is scaled and shifted for display purposes (note the break in the x-axis scale at the lower end). 
    }
    \label{fig:fullselectionscomparisons}
\end{figure*}

Fig.\,\ref{fig:fullselectionscomparisons} compares the two samples in terms of $G$ mag. Both distributions show a well defined peak around $G$\,$\approx$\,16. The drop to fainter mags suggests that the range probed by our sample spans the most interesting range defined by our selection criteria, i.e. we are likely robust to statistical uncertainties at the fainter end. Both samples have significant tails to brighter mags. The mean mags for primary and control are, respectively, $\langle G\rangle$\,=\,15.4 and $\langle G'\rangle$\,=\,15.7. Here, and hereafter, a prime ($'$) superscript refers to the control sample. A difference in mean mags may be expected if excess astrometric noise detection is more efficient for brighter sources. 

The main features that stand out in this plot are the excess of sources in two clumps around $G$\,$\approx$\,11 and 13 in the main sample. The origin of such features is discussed in \citet{lindegren18}, and relates primarily to the `\dof\ bug' identified post-DR2 processing. These features stand out clearly in the distributions of the $\chi^2$ goodness of fit of the astrometric solutions (Ibid.) and, equivalently, in the unit-weight error (\uwe) defined as the $\chi^2_\nu$ with $n$--5 degrees of freedom (see Fig.\,\ref{fig:fullselectionscomparisons}b), where $n$ is the number of good astrometric observations that inform the fit.  Acceptable single-object astrometric fit solutions have an expectation value of $\widehat{{\rm \tt  UWE}}$\,=\,1. For our study, large values of \uwe\ could be indicative of intrinsic complexity beyond a single non-variable source. While aforementioned bug-related artefacts certainly complicate exact comparisons between the primary and control samples, their primary effect is to underestimate the intrinsic noise associated with brighter sources, as discussed in \citet{lindegren18}. Therefore, our selection would, if anything, likely be conservative in terms of \noise\ selection. 

With this selection, the primary sample ends up with a mean excess noise value of $\langle \epsilon \rangle$\,=\,0.60$_{-0.34}^{+0.78}$\,mas, with the uncertainty here representing the 1-$\sigma$ scatter (Fig.\,\ref{fig:fullselectionscomparisons}c). By contrast, the vast majority (99.9\,\%) of control sample objects have \noise\,=\,0 mas,  with just a few showing  \noise\ of up to 0.01\,mas.    

\begin{figure*}
    \includegraphics[angle=0,width=0.99\columnwidth]{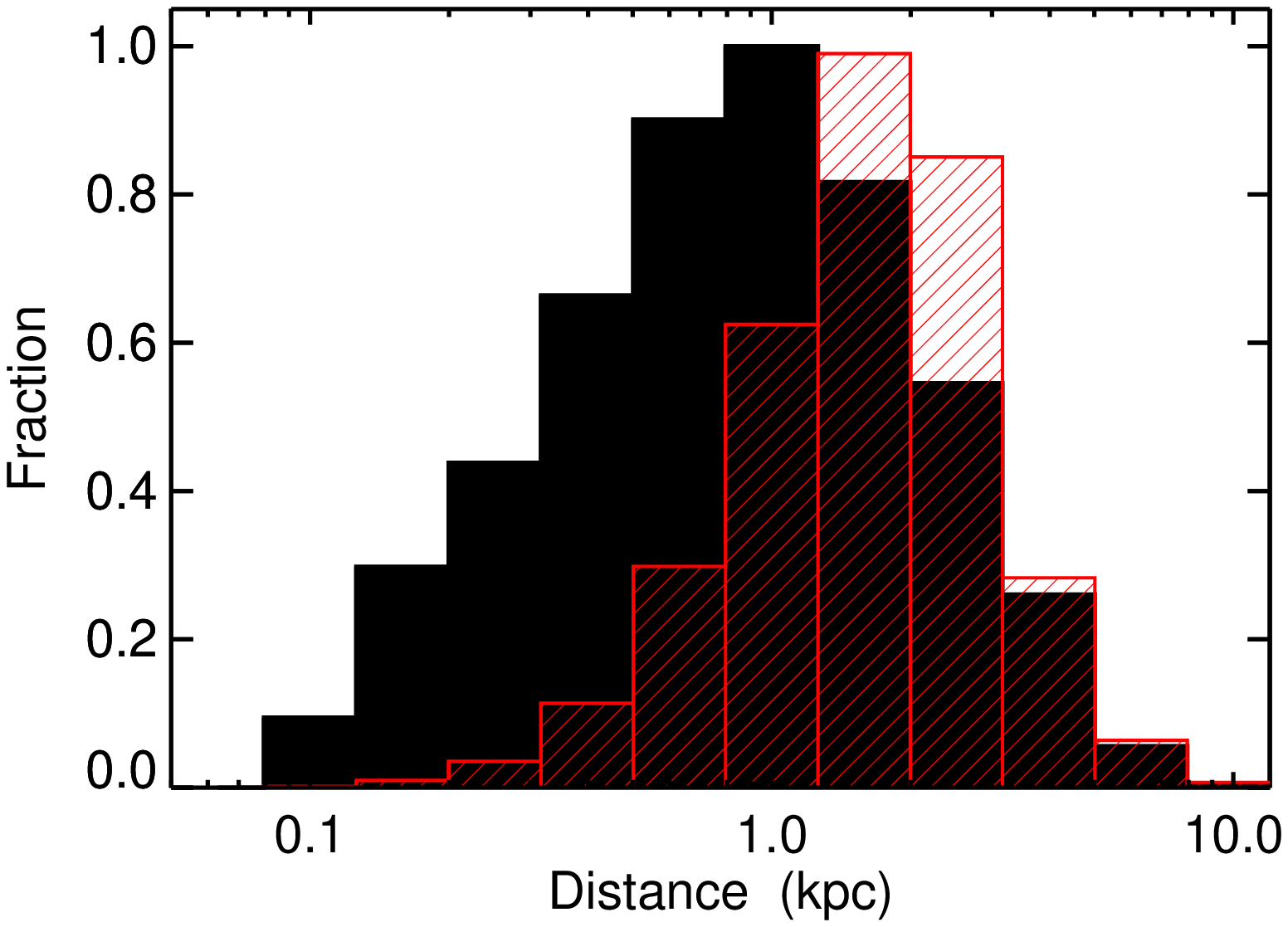}
    \hfill
    \includegraphics[angle=0,width=0.99\columnwidth]{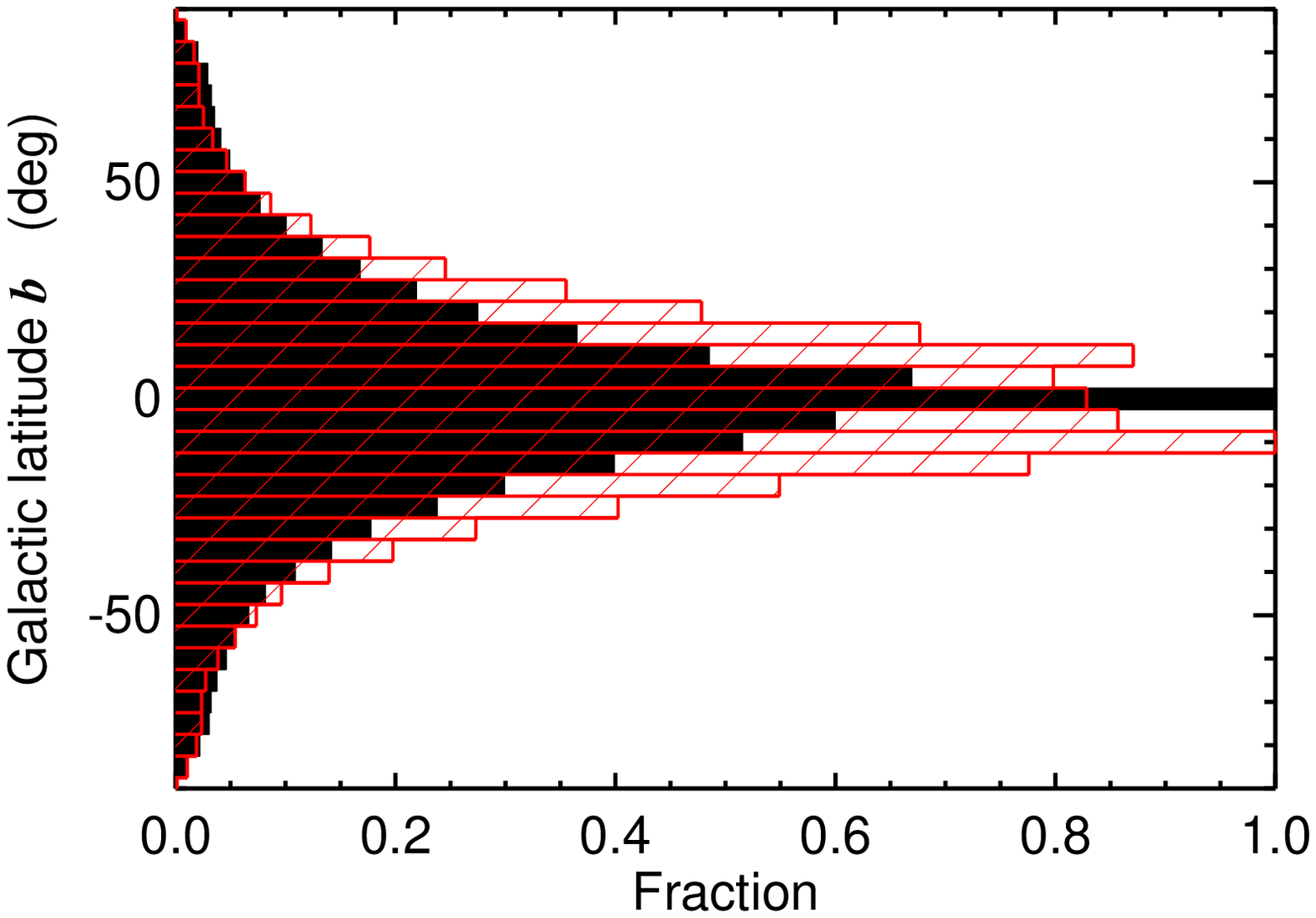}
	\includegraphics[angle=0,width=0.99\columnwidth]{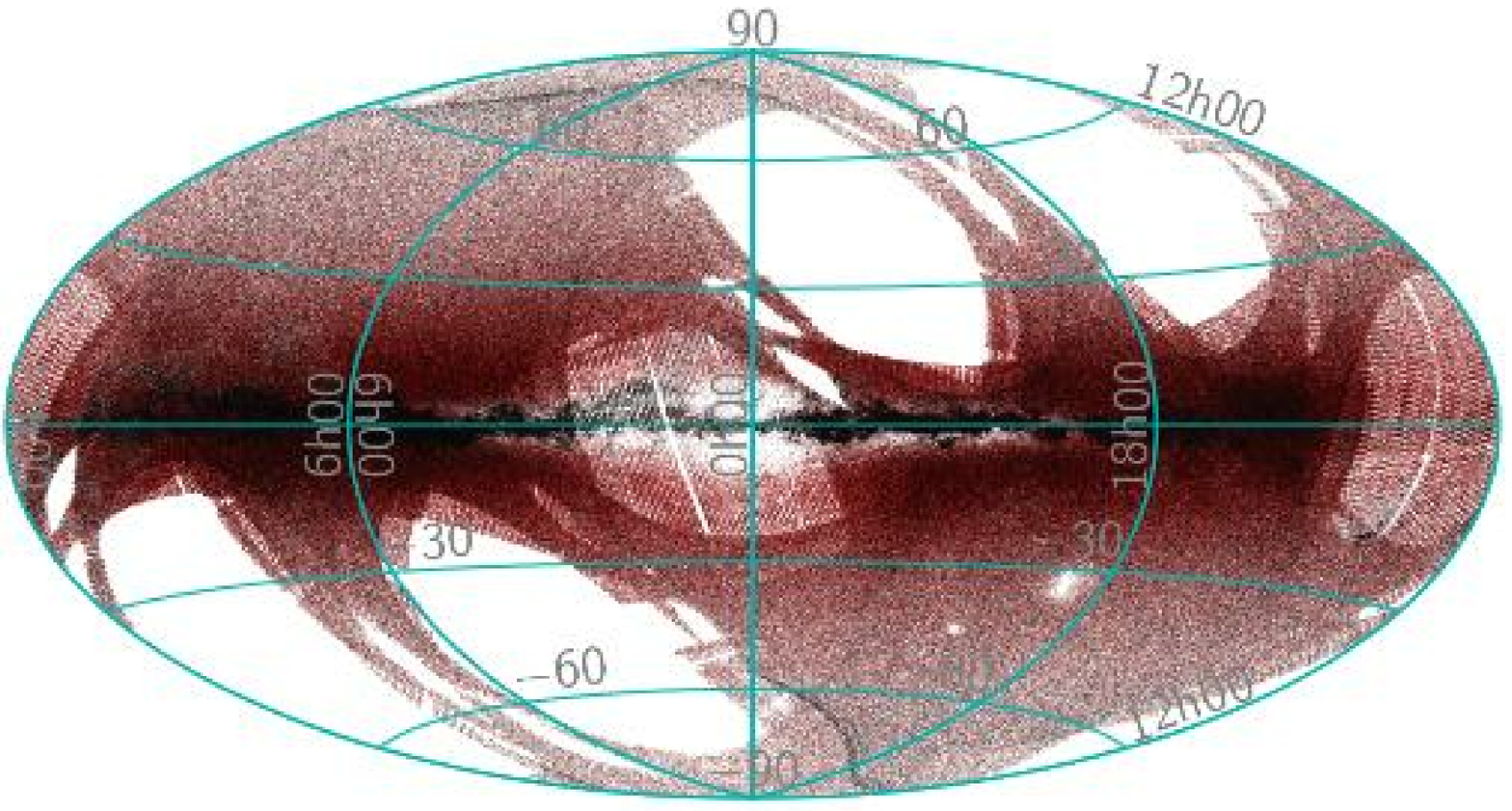} 
    \hfill
    \includegraphics[angle=0,width=0.99\columnwidth]{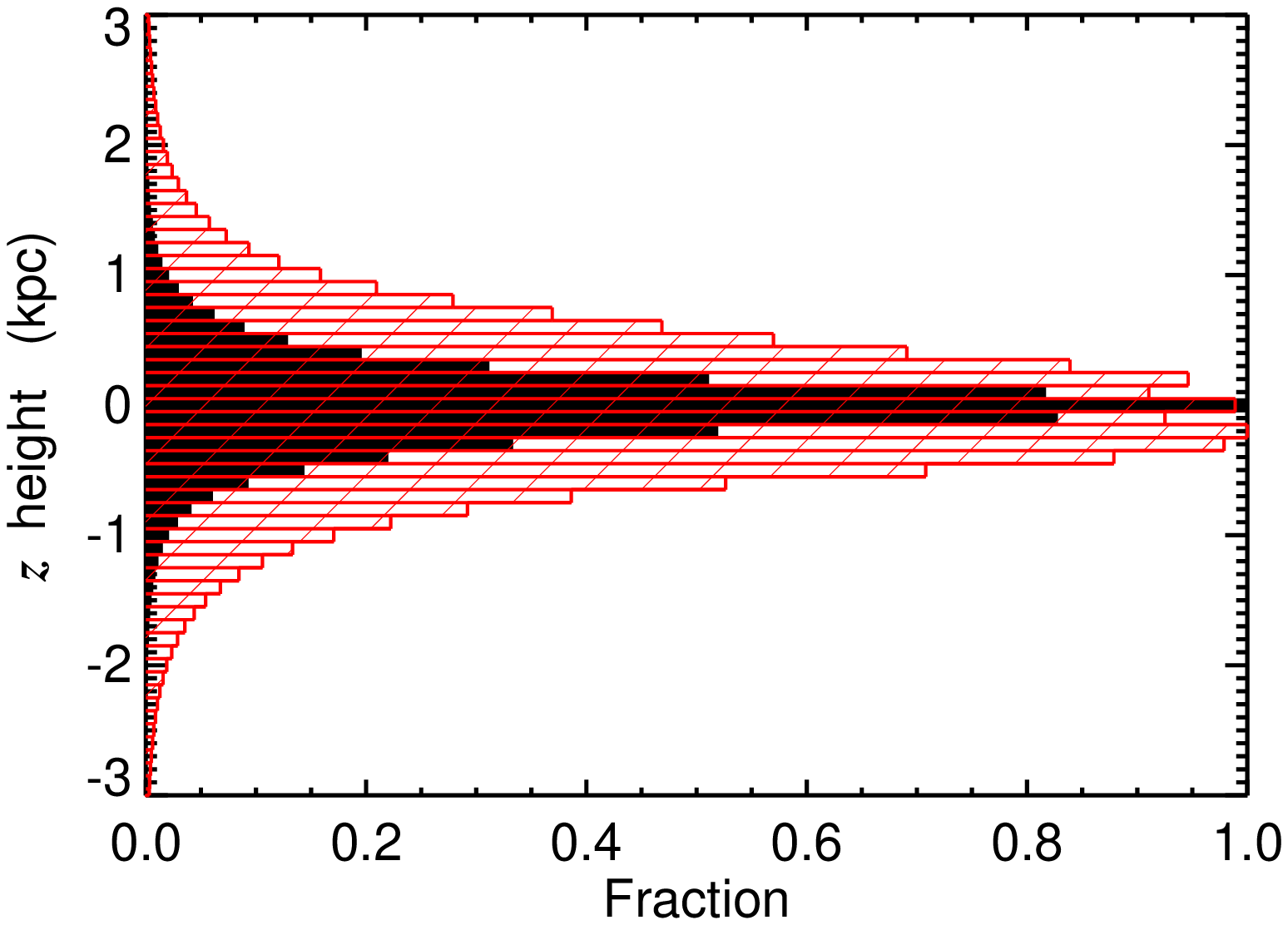}
    \caption{{\bf (a)} Distance distributions of the primary and control samples from Earth;
    {\bf (b)} Histogram of Galactic latitudes $b$; {\bf (c)} Two-dimensional Aitoff sky projection showing one million randomly selected objects from the two samples in Galactic coordinates; 
    {\bf (d)} Corresponding histogram of Galactic latitudinal height $z$ above/below the plane.}
    \label{fig:fullselectionsspatial}
\end{figure*}

The spatial distribution plots in Fig.\,\ref{fig:fullselectionsspatial} show differences that are likely to the intrinsic to the samples. Panel (a) shows the distributions of distances from Earth. Here, distance $d$ is computed simply as $d=\frac{1}{\pi}$, where $\pi$\ is the reported DR2 parallax. Parallax inversion should be a fair estimator of the distance if $\pi$ is measured significantly, and certainly reasonable for population-wide comparisons.\footnote{For our samples, we have 5-$\sigma$, or better, parallax measurements for 87\% and 89\% of the primary and control samples, respectively}. The control sample objects tend to lie farther than primary sources, with mean distances of $\langle d'\rangle$\,=\,1.9\,kpc and  $\langle d\rangle$\,=\,1.2\,kpc, respectively, and a standard deviation of 1.1\,kpc for both. This likely reflects the effectiveness of detecting significant intrinsic perturbations to static single-object astrometric fits (e.g. due to orbital wobble) when sources are nearer. 

Panel (b) of the same figure is arguably more surprising. Here, a difference in the Galactic latitude distribution of the samples is revealed, with the primary sample closely hugging the Galactic plane $\langle b\rangle$\,=\,--0.59$_{-3.9}^{+7.5}$\,deg as opposed to control ($\langle b'\rangle$\,=\,--1.1\,$\pm$\,19.9\,deg). Here, the uncertainties refer to latitudes where the distribution falls of 50\% of its peak, i.e. an effective full-width at half-maximum. It is clear that the primary sample is significantly peakier around the plane. The fractions of primary sample sources within $|b|$\,=\,1, 5, 10 and 20\,deg of the plane are 0.07, 0.25, 0.42 and 0.65, respectively. By contrast, the same fractions for the control sample are 0.04, 0.17, 0.36 and 0.67, respectively. In short, sources with significant \noise\ tend to prefer the plane, as opposed to sources without. 

This difference is marginally evident in the two-dimensional sky distribution in panel (c). Both samples suffer from empty regions devoid of sources, where the \gaia\ sampling (or crowding) does not satisfy our selection thresholds.

But the difference between the samples shows up very prominently in panel (d), depicting the height above the Galactic plane in terms of physical distance. The scatter of the control sample around the plane ($\sigma_{z}^{'}$\,=\,0.54\,kpc) is significantly larger than that for the primary sample ($\sigma_{z}$\,=\,0.41\,kpc). At any given height $z$, the relative fraction of control sample sources exceeds the corresponding fraction of primary sample objects by factors of several. We will return to the interpretation of this distribution in the Discussion section.

\subsection{First insights from colour and a priori classifications}
\label{sec:types}

\begin{figure*}
    \includegraphics[angle=0,width=11.25cm]{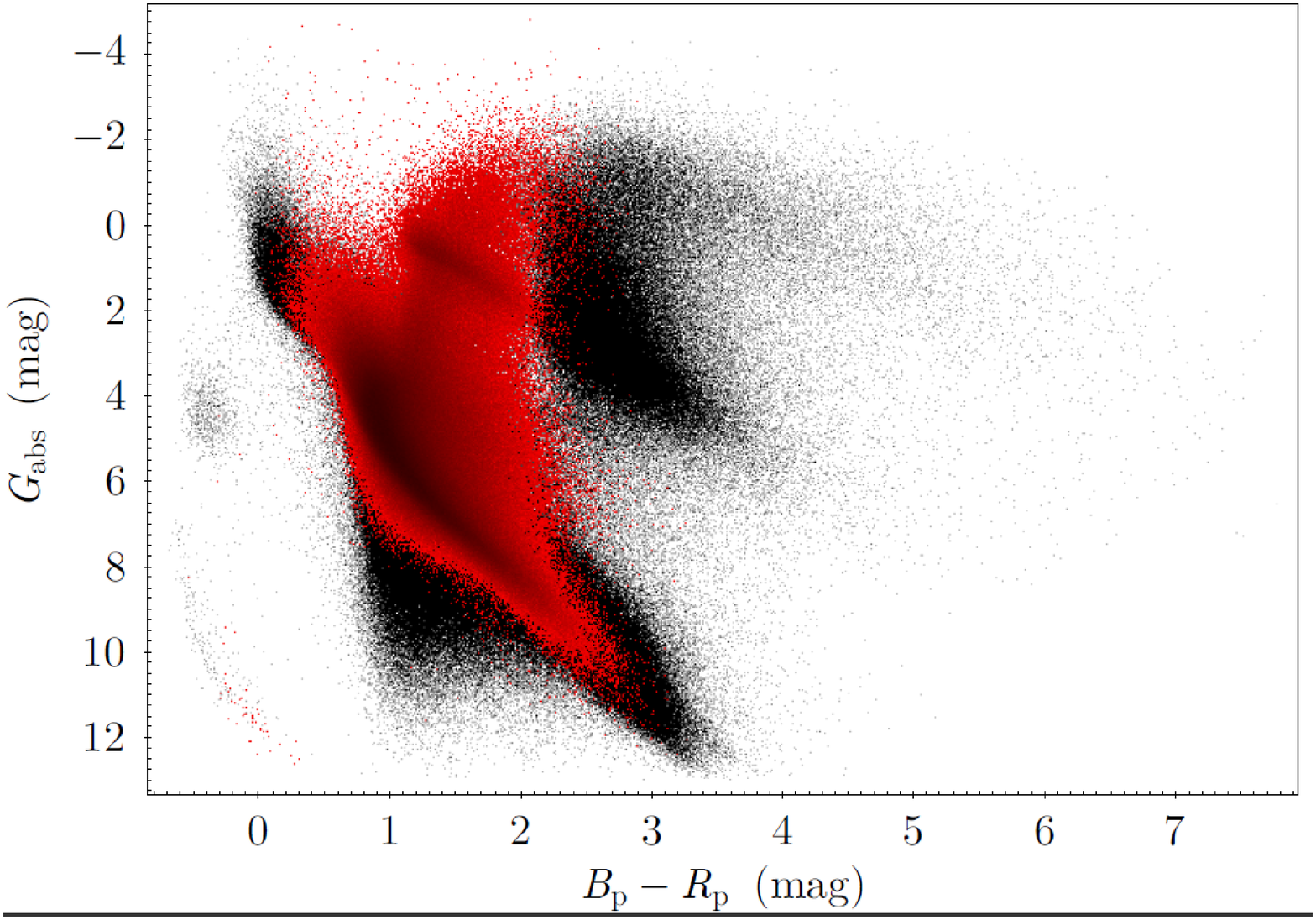} 
    \hspace*{-0.5cm}\includegraphics[angle=0,width=14cm]{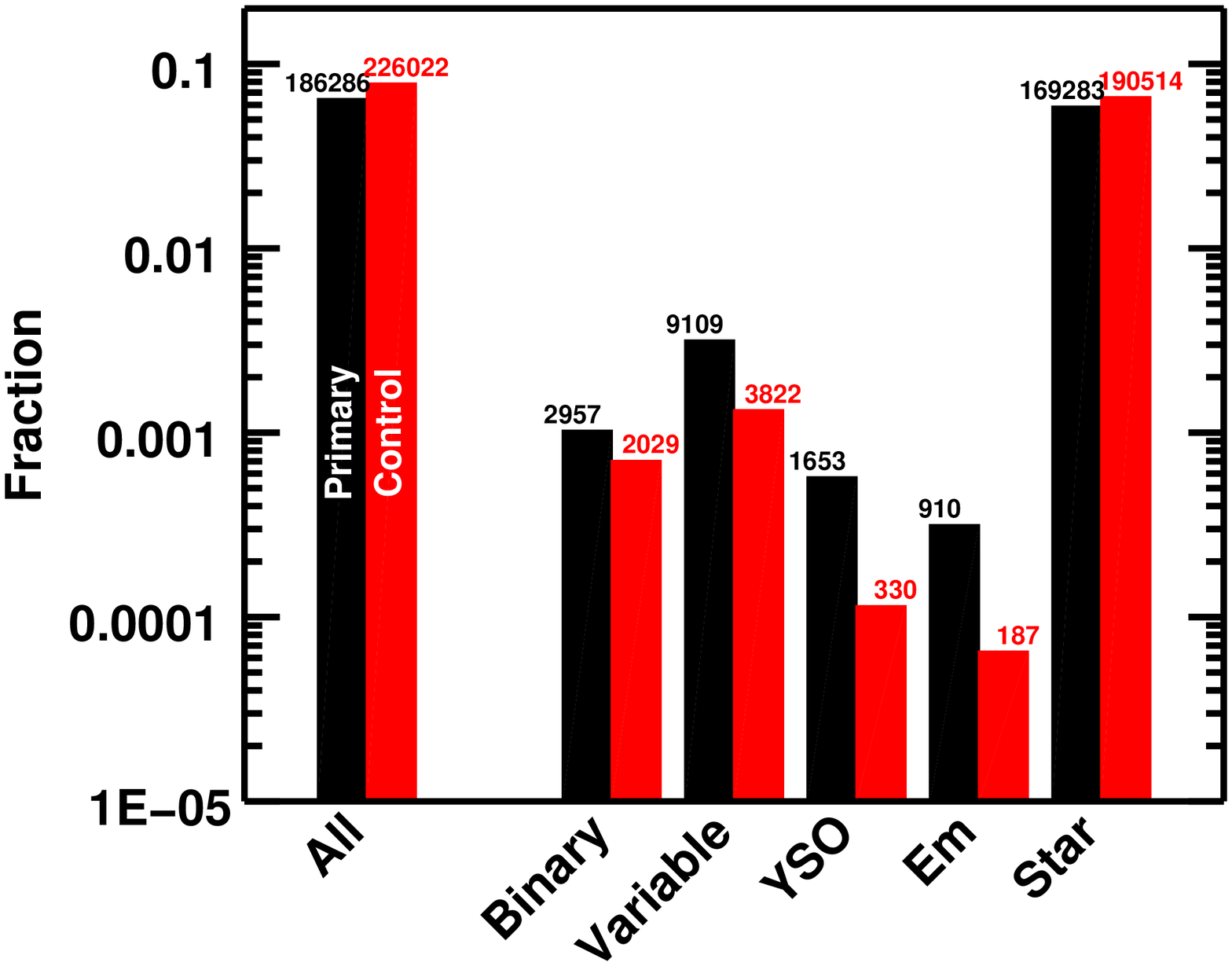}
    \caption{{\bf (a)} The \gaia\ colour vs. absolute magnitude diagram of one million randomly selected objects. 
    No correction for reddening and extinction is applied here. {\bf (b)} Distribution of object classes cross-matched with {\tt SIMBAD}. Here, distance-matching has been applied for a fair cross-comparison. Only a few broad categories are shown (for the full list, we refer the reader to the Appendix). Above each histogram bar, the number of objects in that bin is stated.}
    \label{fig:typeshistall}
\end{figure*}

Fig.\,\ref{fig:typeshistall}a shows the \gaia\ Hertzsprung-Russell (HR) diagram. The control sample's distribution is relatively narrow and concentrated around the main sequence and the giant branch. By contrast, astrometric excess noise selection branches out much more. In particular, the colour of sources with high \noise\ that can be very red. This difference in colour between the samples could arise as a rest of differences in reddening, spectral type, or evolutionary phases being probed. The primary sample also lies at fainter absolute $G$ mag, reflecting their closer mean distances. A small fraction of post-giant branch objects are also apparent in both samples, including the regime occupied by white dwarfs. 

Information regarding known object classifications was collated using a simple cross-match with {\tt SIMBAD}\footnote{\url{http://simbad.u-strasbg.fr/simbad/}}. The `{\tt main\_type}' of the closest association within a threshold distance of 2\,arcsec of the \gaia\ DR2 coordinates was extracted. These were then grouped into a few broad categories for summarisation. The complete list of classes and categories can be found in the Appendix. Sky coverage of classifications is patchy and highly incomplete. But this exercise is solely meant to provide first insight into the putative nature of the population. 

Of the primary sample, 6.8\%\ had a stated classification in {\tt SIMBAD}\footnote{as of August, 2020.}, albeit uncertain in many cases. The corresponding percentage for the control sample was 3.1\%. Much of this difference may be attributable to the differing mean distances of the two samples. It is to be expected that completeness of object type determination will be distance-dependent, with nearby samples being easier to classify. A fairer comparison can be done by distance-matching the two samples. For this, we randomly select one control sample source for every primary sample source, to within a distance threshold of 0.05\,kpc, so that the distribution of distances becomes statistically indistinguishable. The resultant distance-matched sub-sample comprised 2,885,782 objects, and the corresponding distribution of object classes is shown in Fig.\,\ref{fig:typeshistall}b. 

This aligns the two samples much better in terms of the fractions of objects with a stated classification (`All' in the figure). The vast majority of these (85--90\%) have classifications denoting them to be normal stars or similar. There is no substantial difference between the fraction of sources classified as stars, though this classification is potentially more generic than others. In contrast, a significantly higher fraction of objects [$f_{\rm binaries}$\,=\,9.2\,($\pm$\,0.2)\,$\times$\,10$^{-4}$] are classed as binaries, as compared to control [$f_{\rm binaries}'$\,=\,6.7\,($\pm$\,0.2)\,$\times$\,10$^{-4}$]. These include  non-interacting systems as well as interacting binaries such as XRBs and CVs. Three other source types are highlighted here -- variables, emission line objects, and young stellar objects -- as these will be relevant to the Discussion later. In all these cases, the corresponding fraction of systems in the primary sample is significantly higher. See Table\,\ref{tab:fractions} for the listing of the respective fractions. 

\begin{table}
    \centering
    {\tt SIMBAD} object type distributions
    \begin{tabular}{lcc}
\hline
\hline
Class & $f_{\rm Primary}$ & $f_{\rm Control}$\\
      &  10$^{-4}$        & 10$^{-4}$\\
      \hline
    All$^\dag$& 676.6\,$\pm$\,1.5   & 310.6\,$\pm$\,0.3 \\
    &&\\
    Binary& 9.2\,$\pm$\,0.2    & 6.7\,$\pm$\,0.2\\
    Variable& 31.48\,$\pm$\,0.01 & 13.11\,$\pm$\,0.02 \\
    YSO& 4.5\,$\pm$\,0.1    & 0.9\,$\pm$\,0.1\\
    Emission line& 3.2\,$\pm$\,0.1    & 0.65\,$\pm$\,0.05\\
    WR & 0.024$_{-0.009}^{+0.013}$  & 0.007$_{-0.005}^{+0.010}$ \\
    Stars& 489.9\,$\pm$\,1.5    & 661.1\,$\pm$\,1.6\\
    \hline
    \end{tabular}
    \caption{Fractions of object types from {\tt SIMBAD}, split into a few of the key broad categories. $^\dag$Except for the `All' row, the stated fractions refer to the distance-matched \gaia\ (G1+G2+G3) samples in order to enable fair comparisons between primary and control. `All' shows the fractions of classified objects without distance matching, in order to quantify the raw samples.}
    \label{tab:fractions}
\end{table}

\subsection{Multiwavelength properties}
\label{sec:chandra}

The \chandra\ search resulted in 1,490 (4.34$_{-0.11}^{+0.12}$\,\pertenmille{}) 
and 3,222 (0.65\,$\pm$\,0.01\,\pertenmille{}) positive X-ray cross-matches for the primary and the control samples, respectively. The full catalogues are available through CDS\footnote{Reference link to be added upon publication}, and an extract of these is available in the Appendix. We note that not all X-ray detections in the CSC have reported X-ray fluxes and here and hereafter, we only retained the objects where a non-zero broadband X-ray flux measurement is present. Additionally including the objects with zero fluxes would increase the sizes of both samples by $\approx$\,12--15\% (see Table\,\ref{tab:stats}). 

\begin{figure*}
    \includegraphics[angle=0,width=1\columnwidth]{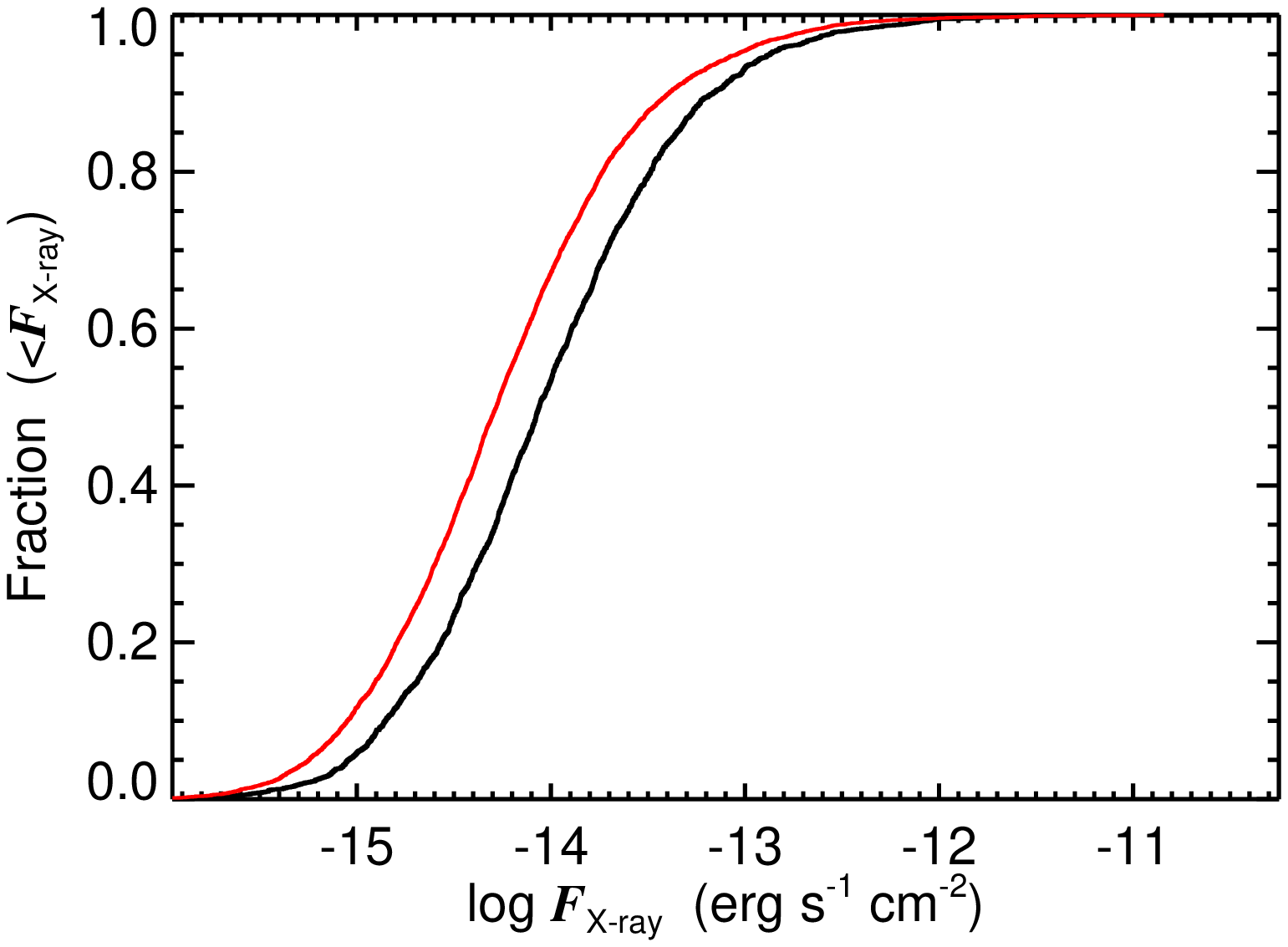}
    \includegraphics[angle=0,width=1\columnwidth]{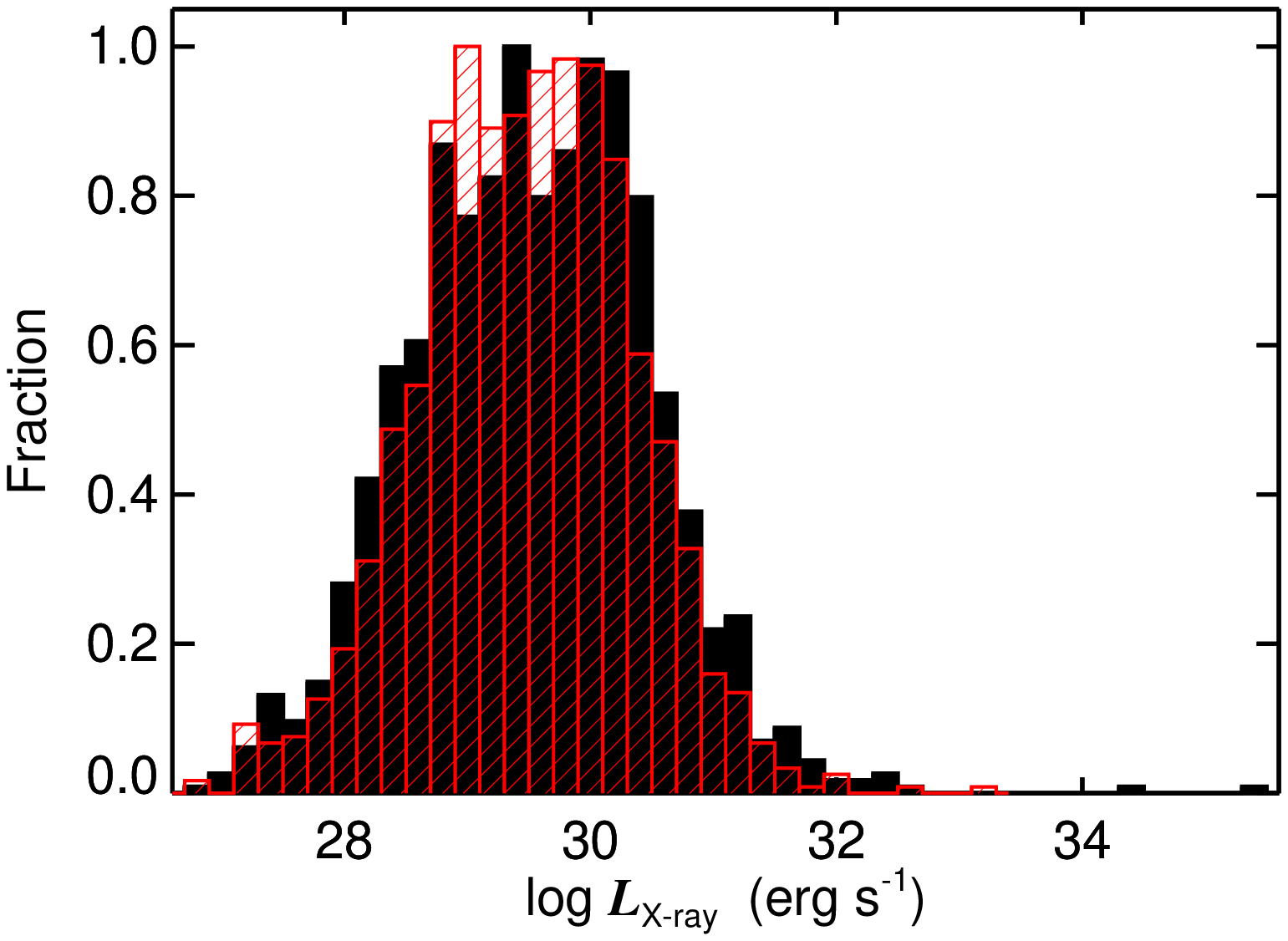}
	\includegraphics[angle=90,width=1.3\columnwidth]{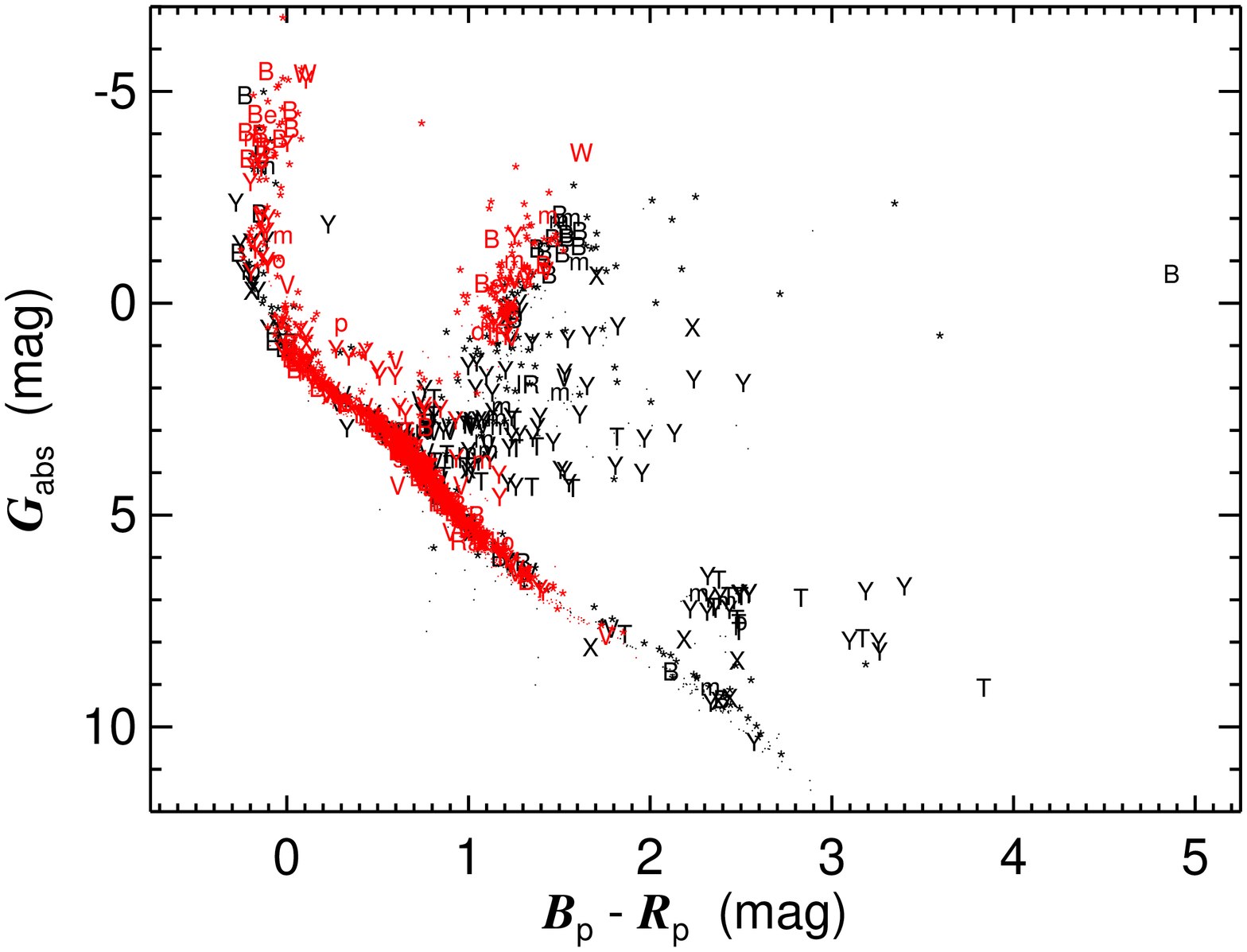} 
    \caption{Comparisons of various properties of the X-ray detected primary and control sample sources. 
    {\bf (a)} X-ray cumulative flux distributions (0.5--7\,keV) of the CSC-matched sources for the main and the control samples. {\bf (b)} X-ray luminosity distributions following distance-matching. {\bf (c)} \gaia\ colour--mag diagram for the X-ray cross-matched samples, with corrections applied for extinction $A_{\rm G}$ and reddening $E(B_{\rm p}-R_{\rm p})$. Individual object types from {\tt SIMBAD} are denoted, with a few key types as follows. `B': Binary, `V': Variable, `Y': YSO, `T': TTau, `X': X-ray source, `W': Wolf-Rayet, `m': Emission line source. The full typing can be found in the Appendix. Objects with a more normal stellar classification (e.g. `Star') are denoted by an asterisk, and those without an archival class are denoted as dots. 
    }
    \label{fig:distributions}
\end{figure*}

Fig.\,\ref{fig:distributions} compares the X-ray matched samples. The  X-ray fluxes of the primary sample tend to be brighter ($\langle F_{\rm X}\rangle$\,=\,6.8\,$\times$\,10$^{-14}$\,erg\,cm$^{-2}$\,s$^{-1}$) than control ($\langle F_{\rm X}'\rangle$\,=\,3.5\,$\times$\,10$^{-14}$\,erg\,cm$^{-2}$\,s$^{-1}$), by a factor of about 2 on average. Here, and hereafter, `X' refers to the broad \chandra\ band of 0.5--7\,keV. 

A comparison of X-ray luminosities requires compensating for the dual dependence on flux as well as distance. A fairer comparison can be done by distance-matching the two samples, as was done for the \gaia--only selection. The mean distances ($d$) of these \gaia+\chandra\ samples now are $\langle d\rangle$\,=\,0.8\,kpc and  $\langle d'\rangle$\,=\,1.1\,kpc, respectively. These are both closer than the respective mean values determined for the \gaia\ only samples because of the additional requirement of X-ray detection. 

The resultant luminosity distribution for the matched samples is shown in Fig.\,\ref{fig:fullselectionscomparisons}d. The distributions are qualitatively similar, peaking close to $\langle {\rm log}[L_{\rm X}\,/\,{\rm erg}\,{\rm s}^{-1}] \rangle$\,=\,29.5. The primary sample displays a scatter ($\sigma_{{\rm log}L_X}$\,=\,1.0) marginally higher than control ($\sigma_{{\rm log}L_X^{'}}$\,=\,0.9), and there is an excess tail of luminous objects in the main sample (e.g. 73 sources [5.3\,$\pm$\,0.7\,\%] with $L_{\rm X}$\,$>$\,10$^{31}$\,erg\,s$^{-1}$, as opposed to 41 sources [3.0$_{-0.5}^{+0.6}$\,\%] for control). We will investigate some of these luminous sources in later sections. 

Some striking differences are visible in the \gaia\ colour--mag diagram for the X-ray cross-matched samples. We now apply extinction and reddening corrections to study the intrinsic source characteristics using the values reported in DR2, where available. We caution, though, and these saturate around $A_{\rm G}$\,$\approx$\,3\,mag and $E(B_{\rm p}-R_{\rm p})$\,$\approx$\,1.5\,mag \citep{andrae18}. Thus, these corrections are underestimated for many individual systems. Nevertheless, canonical features such as the main sequence (MS) and the giant branch immediately stand out. In addition, a clump redder than the MS at relatively faint levels is apparent, corresponding to the expected locus of young stellar objects (YSOs), or to the less-understood population of sub-subgiants \cite[e.g. ][]{geller17}.

The primary sample comprises a significantly higher fraction of evolved and reddened sources than the control sample. Considering the large parameter space corresponding to putative post-MS objects with \bprp\,$>$\,0.9\,mag and \gabs\,$<$\,+4.5\,mag, 32$_{-2}^{+3}$\,\% of the primary sample occupies this regime, whereas the corresponding percentage for the control sample is just 11\,$\pm$\,1\,\%. The scatter of primary sample sources within this region is also larger, suggesting a wide spread of evolutionary phases and source classes, much more so than for control. A further 6\,$\pm$\,1\,\% of the primary sample is found in the putative YSO/sub-subgiant branch regime, whereas no control sources lie in this regime. Finally, the spread of objects is also tighter in the control sample, with the control MS containing no objects fainter than $G_{\rm abs}$\,$>$\,8.4, whereas the primary sample extends about 3\,mags deeper. The source classes are denoted on the diagram, where available. We will delve into these further in the following section. 

We note that a small additional fraction of objects (0.7$_{-0.2}^{+0.3}$\,\%) have unexpected classifications (e.g. extragalactic sources, extended objects such as planetary nebulae [PN], and even candidate planets). Examining the apparent extragalactic sources suggests prior source types are probably in error (either simple transcribing errors between {\tt SIMBAD} and published work (cf. Appendix), or source confusion. All of them have significant positive parallax measurements consistent with being Galactic objects. The reason that a few PN lie in our primary sample remains unclear; e.g. whether or not the extended nebular emission introduces artificial astrometric uncertainties. Such objects should obviously be treated with caution; but given their small numbers, they will not bias any of the inferences below.

\begin{table*}
Luminous X-ray sources in our primary sample based upon astrometric excess noise\\
\begin{tabular}{lcccccr}
\hline
\hline
Candidate & RA  &  Dec  & $L_{X}$ & $d$ & $\epsilon$ & Name (Type)\\
number & deg  &  deg  & 10$^{32}$\,erg\,s$^{-1}$ & kpc & mas & \\
\hline
   1 &   254.45740 &    35.34251 &  3052.5 & 6.72\,$\pm$\,1.20& 0.21\,$\pm$\,0.02 & Her\,X--1 (XRB)\\ 
   2 &   152.44570 & --58.29341 &   209.8 & 4.10\,$\pm$\,0.55& 0.25\,$\pm$\,0.02 & GRO\,J1008-57 (XRB)\\
   3 &   156.07670 & --57.80831 &     3.2 & 3.57\,$\pm$\,0.57& 0.13\,$\pm$\,0.02 & WR\,20b (WR)\\
   4 &   206.63570 & --62.92333 &     3.1 & 1.69\,$\pm$\,0.13& 0.14\,$\pm$\,0.02 & HD\,119682 (Be*) \\
   5 &   161.56250 & --60.42921 &     2.6 & 4.60\,$\pm$\,0.64& 0.23\,$\pm$\,0.02 & UCAC2\,5304815 (Em*)\\
   6 &   260.13240 & --35.85320 &     2.1 & 1.61\,$\pm$\,0.17& 0.15\,$\pm$\,0.02 & 2E\,1717.1-3548 (Candidate OB*) \\
   7 &    26.86804 &    61.80746 &     2.0 & 6.21\,$\pm$\,1.39& 0.15\,$\pm$\,0.02 & VES\,629 (Em*) \\
   8 &   308.17060 &    41.24152 &     1.7 & 0.85\,$\pm$\,0.09& 0.59\,$\pm$\,0.00 & Schulte\,12 (Blue\,SG) \\
   9 &    53.74959 &    53.17319 &     1.3 & 6.98\,$\pm$\,1.74& 0.20\,$\pm$\,0.02 & V0332+52 (XRB) \\
\hline
\end{tabular}
\caption{List of \gaia--CSC matched sources with $L_{\rm X}$\,$>$\,10$^{32}$\,erg\,s$^{-1}$ (0.5-7 keV)\label{tab:lum}}
\end{table*}

\section{Discussion}
\label{sec:discussion}

We have selected a sample of Galactic point sources with significant astrometric excess noise (\noise) that are also X-ray detected, with the aim to explore the possibility of detecting previously unidentified or ill characterised interacting binaries. In order to better comprehend the nature of \noise-based selection, we also collated a control sample with complementary low values of \noise. A \gaia\ selection of isolated (within 5\arcsec) objects and clean photometry results in about 3\,million sources with significantly detected \noise\ (together with significant parallaxes for distance estimation). A control sample is defined to have low complementary values of \noise, but with other selection criteria being identical; this is about 14 times more abundant than the primary sample (Table\,\ref{tab:stats}).

\subsection{Summary of observed differences between primary and control}

The following comparisons between our primary and control samples are revealing: 

\begin{enumerate}
    
    \item Sources with significant \noise\ lie closer to us than control sources ($\langle d'\rangle$\,=\,1.9\,kpc and $\langle d\rangle$\,=\,1.2\,kpc; $\S$\,\ref{sec:fullsample} and Fig.\,\ref{fig:fullselectionsspatial}a).
    
    \item The primary sample is distributed closer to the Galactic plane, on average. The overwhelming majority ($>$\,98\,\%) of sources lie within 1\,kpc of the Galactic plane (in terms of height $z$). When probing even closer in, the percentages of sources within $|z|$\,$<$\,0.1\,kpc is 68\,$\pm$\,3\,\% for primary and 59\,$\pm$\,2\,\% for control, respectively. This preference for the primary sample reflects its aforementioned preference for low Galactic latitudes $b$. 
    
    \item Published object classifications are available for only a small percentage of the sample ($\sim$3--6\,\%). After matching the two samples to account for the fact that object classification may be distance-dependent, we find a higher percentage of objects classified as binaries, variables, emission line sources and young stellar objects in the primary sample, as opposed to control ($\S$\,\ref{sec:types} and Fig.\,\ref{fig:typeshistall}). A wide variety of detailed object classes fall under these categories, and the classifications also include candidate and uncertain types. Nevertheless, these caveats should not affect the relative comparison between the primary and the control samples herein. 
    
    \item The X-ray detection fraction with \chandra\ in our primary sample is significantly higher (about 7 times more) than control. There is no dramatic difference in the distance-matched comparison of X-ray luminosities of the samples, except for a excess tail of luminous objects in the primary sample ($\S$\,\ref{sec:chandra}). The samples are able to probe down to the level where quiescent emission from XRBs, active stars and YSOs can be detected. 
    
    \item Occupancy in colour--mag space across the HR diagram differs between the two samples, with the primary sample extending to lower mags than control, and showing a wider spread away from the main sequence ($\S$\,\ref{sec:chandra} and Fig.\,\ref{fig:distributions}e). 

\end{enumerate}

\noindent
The nature of \noise\ selection, its value and its veracity still remain unclear, since DR2 is an intermediate data release, and the individual astrometric measurements have not yet been released. This is why we chose to explore the influence of \noise\ selection in a controlled manner and in a relative sense between two closely similar samples. 
While any one of the above differences may be attributable to selection effects or systematic uncertainties in astrometric fitting, such explanations do not suffice when considering the above differences cumulatively. In particular, systematic effects in any one mission (e.g. \gaia/DR2) or wavelength would not obviously be expected to translate into differences at other wavelengths (\chandra) or catalogues ({\tt SIMBAD} spectroscopic classifications). Thus, the above differences suggest that \noise\ selection is likely effective in picking up sources with intrinsically distinct properties, on average. 

\subsection{Combining Optical and X-ray photometry}

\begin{figure}
    \includegraphics[angle=0,width=1\columnwidth]{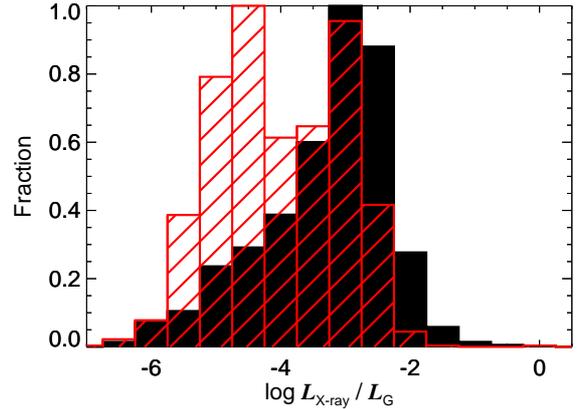}
    \caption{The X-ray to $G$ band luminosity ratio, as a proxy for the X-ray bolometric fraction. The primary sample (black) shows a mean ratio higher than control (red) by $\approx$\,0.6\,dex.}
    \label{fig:lxlg}
\end{figure}

One can take the above tests further by combining data at multiple wavelengths. In Fig.\,\ref{fig:lxlg}, we compare the optical against X-ray properties of the two samples. The X-ray to Bolometric (Bol) flux ratio for a wide variety of stellar classes (including chromospheric and coronal activity, shocks in winds and ejecta) saturates around 10$^{-4}$ to 10$^{-3}$ \citep[e.g. ][]{vilhuwalter87, testa10}. 

The figure reveals a clear separation in the average luminosity ratio $l_{XG}$\,=\,log($L_{\rm X}$/$L_{\rm G}$), with $\langle l_{XG}\rangle$\,=\,--3.3 and $\langle l_{XG}^{'}\rangle$\,=\,--4.0. The difference is highly significant based upon a KS test. The scatter for both samples is $\approx$\,0.9 dex, with the tail to lower luminosities being more extended for control. Here, for simplicity, we have used the $G$ band \gaia\ luminosity as a proxy for $L_{\rm Bol}$, so the absolute values are less relevant than their relative distributions. Nevertheless, the difference is consistent with intrinsic distinctions between the two samples. 

$l_{XG}$ ratios higher than the saturation threshold above are indicative of more efficient X-ray generation processes such as accretion, though this is not necessarily a unique solution and YSOs are also known to be copious producers of X-ray emission \citep[e.g. ][]{feigelson99}. Thus, more detailed investigations of individual sources are warranted. 

We also caution the reader that the $G$ band luminosities may be underestimated due to the effects of dust reddening, which could be underestimated in cases where the \gaia\ extinction estimates saturate (see \S\,\ref{sec:chandra}). In order to account for the distance-dependent influences of reddening through the Galactic plane, we carried out the same exercise above for our distance-matched main and control samples. This yielded almost identical results as above in terms of the mean luminosity ratios, albeit on smaller subsamples. 

We also took this test further by matching the primary and control samples on {\em extinction} $A_{\rm G}$. A threshold of $\Delta A_{\rm G}$\,=\,0.01\,mag was used was matching every primary sample source to a corresponding control sample object. Thereafter, objects with zero X-ray flux were removed. This resulted in 417 and 412 objects in the extinction-matched primary and control subsamples, respectively, with mean values $\langle A_{G}\rangle$\,=\,0.9\,mag for both. The distribution of extinctions was confirmed with be indistinguishable based upon a KS test. For these subsamples, we find $\langle l_{XG}\rangle$\,=\,--3.7 and $\langle l_{XG}^{'}\rangle$\,=\,--3.9. In other words, the trend of higher luminosity ratios in the primary sample is preserved, but with a milder difference with respect to the control sample. 

\subsection{The X-ray luminous subsample}
\label{sec:lum}

Next, we examined a subset of objects, choosing the most luminous X-ray sources as examples to probe individually. Table\,\ref{tab:lum} shows these systems, choosing $L_{\rm X}$\,$>$\,10$^{32}$\,erg\,s$^{-1}$ as a threshold of interest here. 

Optical/infrared counterparts are known in all cases, though detailed classification and follow-up are not necessarily always available. Two things are noteworthy in this sample. Firstly, there is an obvious wide range in the types of objects that we find, from WR systems to Be and other emission line stars, and XRBs. 

Secondly, there are 3 known XRBs (33$_{-21}^{+38}$\,\%) in this short list. This is particularly interesting given that most XRBs tend to lie at distances beyond $\sim$\,1\,kpc (e.g. \citealt{gandhi19}), whereas our sample has $\langle d \rangle$\,=\,0.8\,kpc below 1\,kpc; \S\,\ref{sec:chandra}.

By contrast, the control sample includes 24 objects with $L_{\rm X}$\,$>$\,10$^{32}$\,erg\,s$^{-1}$, 6 of which are classified as XRBs (25$_{-11}^{+16}$\%). The fraction of XRBs is consistent between the two samples. But a fairer comparison would be to measure the fraction amongst the respective distance-matched samples. After distance-matching, the control sample contains only two sources above $L_{\rm X}$\,$>$\,10$^{32}$\,erg\,s$^{-1}$, neither of which is an XRB. At face value, these comparisons are again consistent with an excess of interacting binaries in the primary sample. 

One caveat to keep in mind while examining X-ray luminosities is that the CSC fluxes may not reflect the long-term average (or quiescent) level. It is possible that pointed observations of transient systems were requested and carried out when they were undergoing particularly interesting phases, such as brightening, outbursts or fades. However, such an observation bias, if operational, should affect both our primary and control samples in the same way, so that relative comparisons would be insensitive to it.

\subsection{The magnitude of \noise}
\label{sec:noisemag}

\begin{figure}
    \includegraphics[angle=0,width=1\columnwidth]{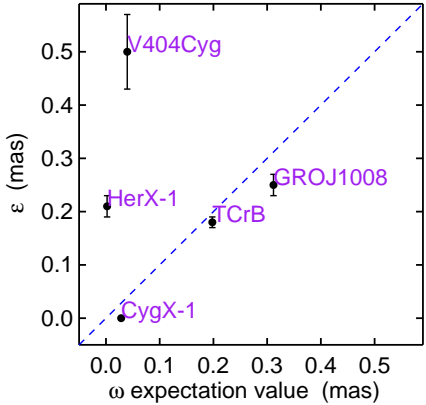}
    \caption{Direct comparison of measured astrometric excess noise (\noise) on the y-axis and the expected value of orbital wobble ($\hat{\omega}$) on the x-axis. The comparison is shown for a few select sources with known orbital parameters. The dashed line is the 1:1 locus.}
    \label{fig:comparewobble}
\end{figure}

Examining the system properties quantitatively, however, yields a more complex picture. This is particularly true when examining the DR2 reported {\em magnitude} of \noise. As discussed in Section\,\ref{sec:understanding}, the expectation value of \noise\ should approximate $\hat{\omega}$, assuming that orbital wobble dominates over any other source of uncertainty in the fitted astrometric solution. Fig.\,\ref{fig:comparewobble} compares these quantities for a handful of well-known interacting binaries for which the system parameters have been measured (or estimated), and consequently, $\hat{\omega}$ can be computed. These include V404\,Cyg (\noise\,=\,0.50\,\p\,0.07\,mas; cf. \citealt{gandhi19}), Cyg\,X--1 (\noise\,=\,0; \citealt{gandhi19}), T\,CrB (\noise\,=\,0.18\,\p\,0.01\,mas; cf. \citealt{schaefer18}) and Her\,X--1 and GRO\,J1008--57 from Table\,\ref{tab:lum}. We note that was excluded by our near neighbour criterion (G2) and T\,CrB. Irrespective, all these systems are known to be prolific X-ray emitters during accretion-driven outbursts. 

\begin{figure}
    \includegraphics[angle=0,width=0.92\columnwidth]{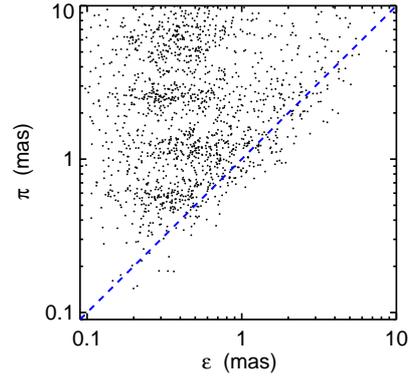} 
    \caption{Parallax ($\pi$) vs. astrometric excess noise (\noise) for the primary X-ray matched sample. The dashed line is the 1:1 locus.}
    \label{fig:wobbleparallax}
\end{figure}

Cases where \noise\ is less than $\hat{\omega}$ can be understood as resulting from the sensitivity  and sparseness of \gaia\ sampling. But objects with \noise\,$\gg$\,$\hat{\omega}$ cannot be explained similarly. This is true for objects such as Her\,X--1 and V404\,Cyg. The former is a persistent NSXB, while the latter is a transient BHXB. \noise\ exceeds $\hat{\omega}$ by factors of 10 and 70 in the two systems, respectively. The presence of a near neighbour (within 2$''$) for V404\,Cyg could be causing deviations from a single-source astrometric fit. Another possibility may be related to the fact that both systems are low-mass XRBs, known to show strong optical flux variability. V404\,Cyg underwent a dramatic outburst in 2015, undergoing variability on a variety of timescales by up to 7\,mag \citep[e.g., ][]{kimura16, g16_v404}. Her\,X--1 is an eclipsing system with known orbital and superorbital flux modulations on a range of timescales \citep{jurua11}. Such variations would introduce systematic variations in the epochwise astrometric uncertainties. But given that the outburst of V404\,Cyg lasted only a few weeks, such changes are not expected to dominate across the full DR2 observation period. 

Therefore, the presence of some additional noise floor is a possibility to be considered. \citet{belokurov20} have studied the distribution of \noise\ and show that its scatter is large for small values of expected centroid shift of $\ltsim$\,0.2\,mas (cf. their Fig.\,A2), in the regime where transient XRB wobbles are expected to lie (cf.\,Fig.\,\ref{fig:wobble}). One instrumental cause for this may be attitude errors \citep[e.g., ][]{lindegren12}, which have been globally 
adjusted in DR2 for weighted residuals to align with DOF \citep{lindegren18}. However, the extent of any cross-talk between the attitude noise and source excess noise terms remains unclear, and future data releases will shed more light on this issue.  

Such adjustments may also explain why source parallaxes do not appear to be dramatically influenced, even in the case of large apparent \noise. A good example of this is, again, V404\,Cyg, whose DR2 parallax $\pi$\,=\,0.44\,$\pm$\,0.10 is {\em smaller} than the reported noise value of \noise\,=\,0.50\,$\pm$\,0.07. Fig.\,\ref{fig:wobbleparallax} compares these two quantities for all of our 1,490 primary sample sources. As expected, \noise\ never substantially exceeds \w, though there are relatively small excesses in many cases. For 141 (9\,$\pm$\,1\% of) sources, \noise\ exceeds $\pi$ as in V404\,Cyg, with the mean excess in these cases being $\langle \frac{\epsilon}{\pi}\rangle$\,=\,1.3.

\subsection{Using \ha\ as an activity indicator}
\label{sec:ha}

\begin{figure*}
    \includegraphics[angle=0,width=0.65\columnwidth]{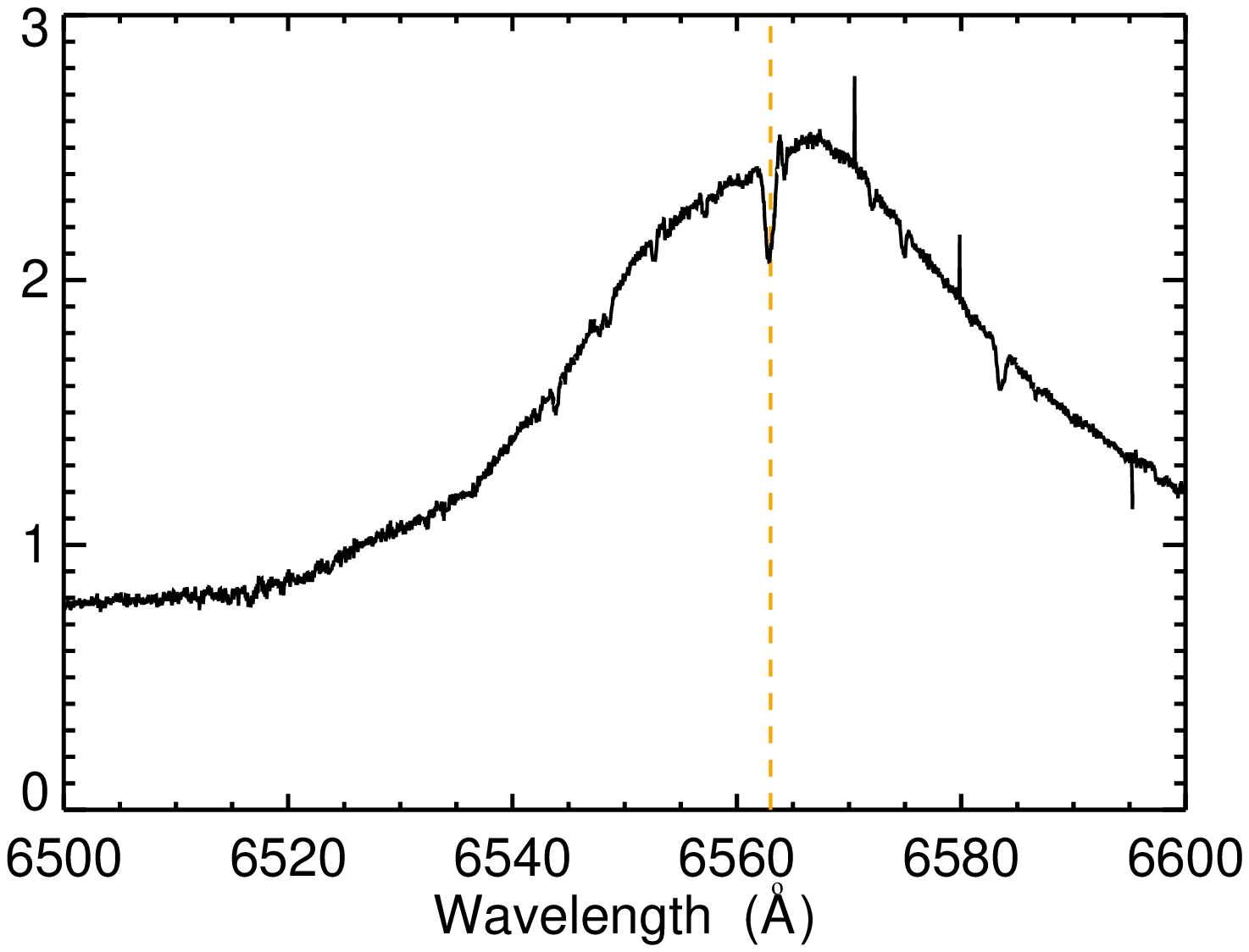}
    \includegraphics[angle=0,width=0.65\columnwidth]{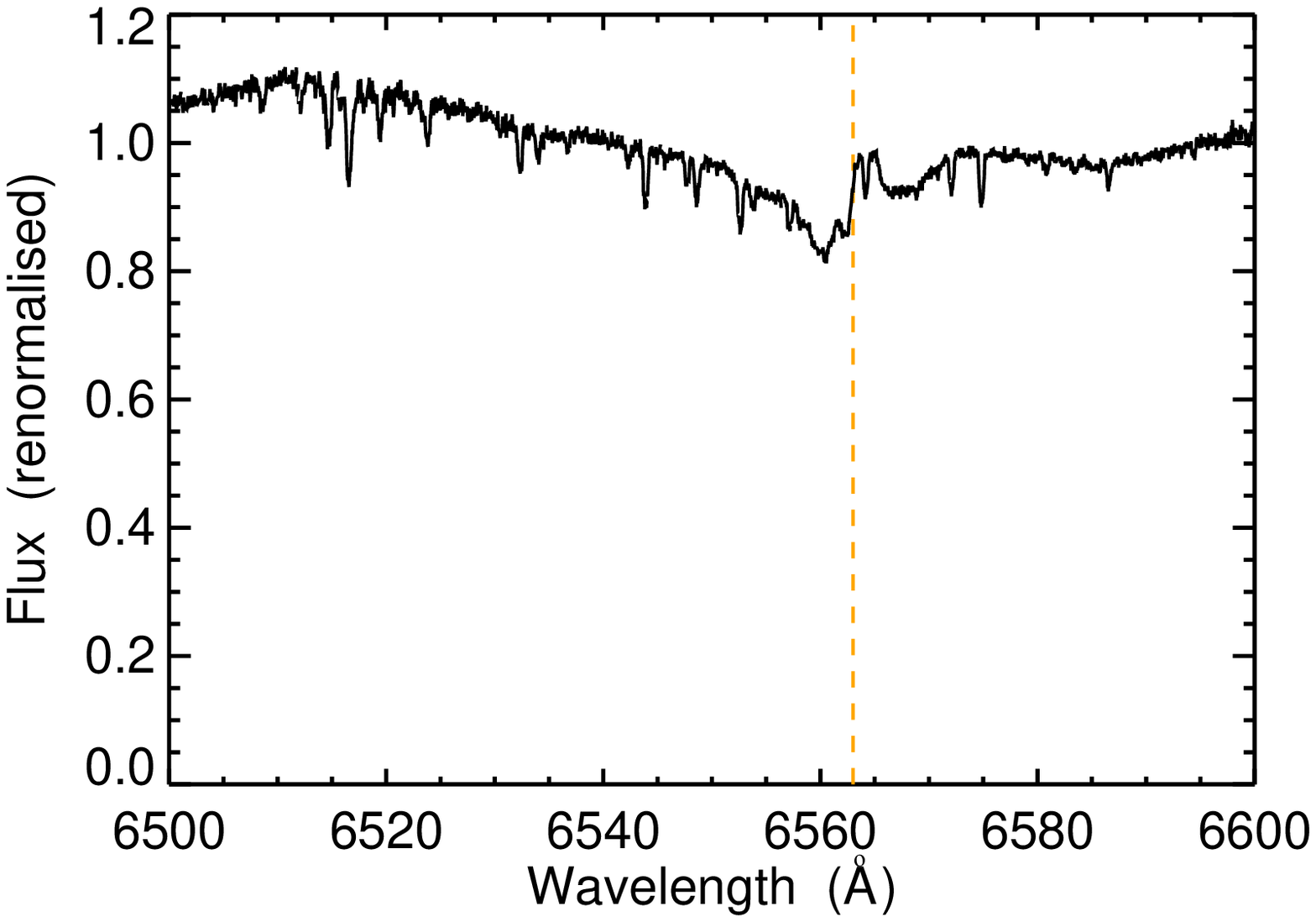}
    \includegraphics[angle=0,width=0.65\columnwidth]{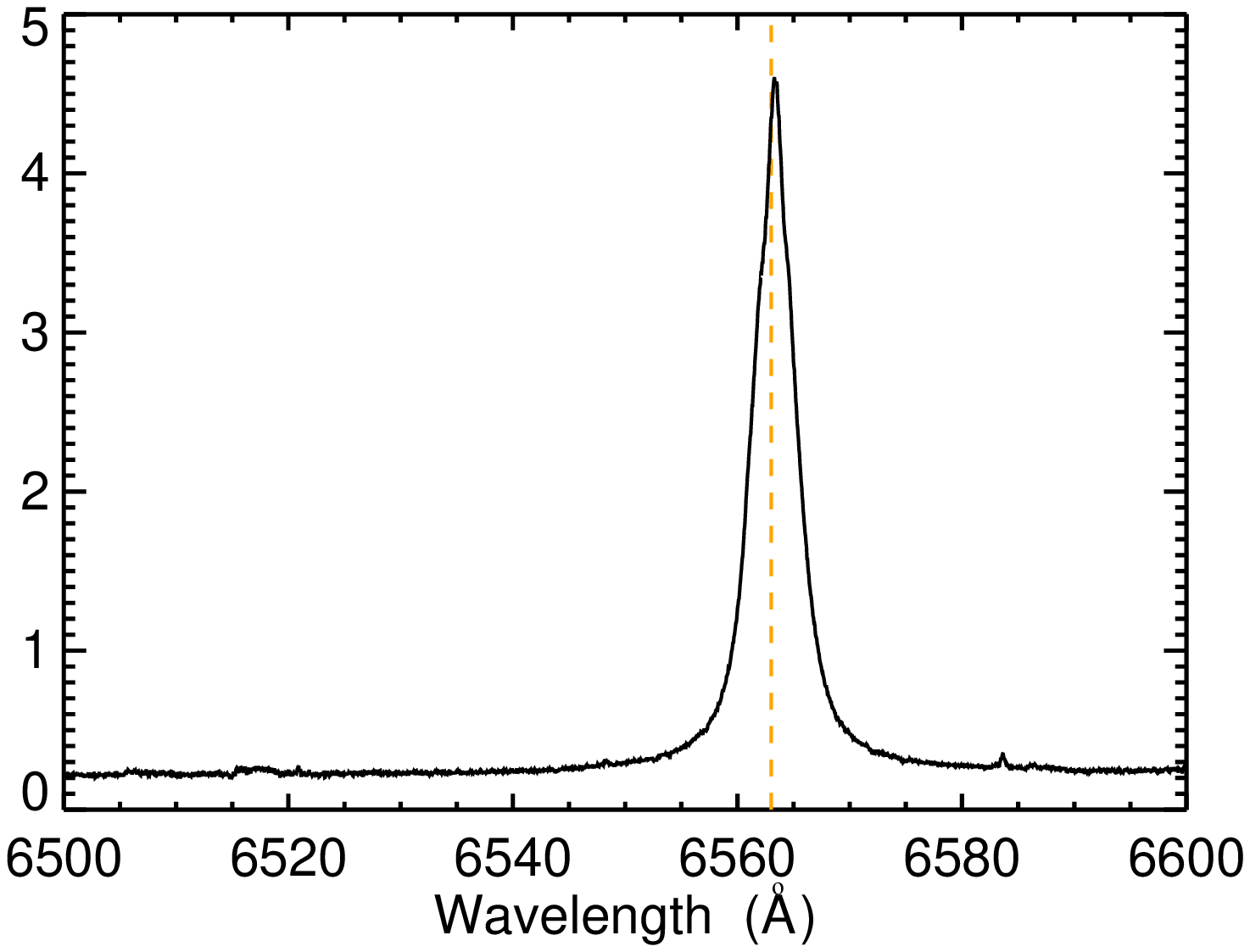}
    \caption{SALT HRS spectrum of sources \#3, 4, 5, respectively, left to right, from our luminous sample listed in Table\,\ref{tab:lum}. The dashed vertical line denotes the wavelength of \ha, showing the presence of a very broad line in \#3, a P\,Cyg profile in \#4, and a narrow line in \#5, respectively. 
    }
    \label{fig:hd119682}
\end{figure*}

The above investigation suggests that caution is required when interpreting the reported DR2 values of \noise. 
To further test the validity of our sample, we investigated the presence of \ha\ emission as an additional indicator of intrinsic activity, for instance as result of the presence of viscously heated accretion discs. In Fig.\,\ref{fig:hd119682}, we first present our SALT follow-up of three objects from our sample of luminous sources, chosen entirely at random from amongst the objects not already known to be binaries, and those that were visible and bright enough for spectroscopy from South Africa in mid-2020. Double-peaked \ha\ emission is present in one object (\#4), together with what appears to be a P-Cygni profile indicative of an outflow. Both the other sources show clear and strong \ha\ in emission. As discussed in \S\,\ref{sec:lum}, source classification is available for some of the objects in our sample, so \ha\ detection is not a surprise for the cases of the WR star (\#3) and the Be system (\#4). However, this classification was not used while making the selection for observation. Moreover, we are unaware of any publication explicitly reporting and showing an  optical spectrum for \#4. Almost nothing is known about \#5 and the nature of its \ha\ emission will need further study. Though the spectra clearly indicative active processes powering emission lines in all these, none of these prove the presence of a binary in any of the systems\footnote{Prior studies on the binary nature remain controversial for \#4, HD\,119682 \citep[e.g., ][]{langer20}, whereas short period binary solutions appear to be ruled out for \#3, WR\,20b \citep[e.g., ][]{skinner10}.}, and radial velocity searches will be needed.

Going beyond individual systems, we next compared the \ha\ detection rate of the primary vs. control populations using IPHAS narrow-band photometry. Of 3,222 objects in the control sample, 292 
systems had a counterpart in the \gaia/IPHAS catalogue, whilst out of the 1490 
systems in the primary sample, 171 
had a counterpart \citep{scaringi18}, both within 1$''$. We then identified \ha-excess sources from these subsets using a catalogue in preparation (Fratta et al.) which uses the combination of \gaia\ and IPHAS to select systems displaying \ha-excess relative to their underlying stellar population based on the calibrated \gaia\ colour--magnitude diagram. Of 292 
objects from the control sample, 20 
(7\,$\pm$\,2\,\%) are found to be \ha-excess sources, whilst of 171 
systems in the primary sample, 31 
systems (18\,$\pm$\,4\,\%) are found to have \ha-excess. The difference between the samples is significant, consistent with a larger fraction of active systems in the primary sample.

These \ha\ studies imply that our primary sample is unlikely to be dominated by false positives, and that our selection is not biased by objects that are undergoing any recent dramatic variability in their spectral types (since \gaia\ began operation). The individual spectra also serve to demonstrate the spectral diversity of systems that astrometric noise selection is capable of picking out. All of these are clearly active in a variety of ways, and we explore the nature of the sample in the next section.

\subsection{The nature of the primary sample and the search for new accreting binaries}

We began this study with the aim of searching for putative new accreting binaries. Are we likely to have such candidates amongst our primary sample? 

The mean value of excess noise across our primary sample, $\langle \epsilon \rangle$\,=\,0.60$_{-0.34}^{+0.78}$\,mas, at the mean distance of 0.9 kpc, suggests that if accreting binary systems dominate our sample, they ought to have properties similar to long period systems such as T\,CrB (cf. Fig.\,\ref{fig:comparewobble}).  Our overall distribution of \noise\ (Fig.\,\ref{fig:fullselectionscomparisons}c) rises below \noise\,=\,1\,mas and peaks around 0.2\,mas, before decreasing again, suggestive of incompleteness below this level. Together with the results from the numerous tests in the previous sections testing the veracity of our sample, we conclude that DR2 {\tt astrometric\_excess\_noise} selection works effectively in an average sense, but does not match the expectations for individual sources quantitatively, especially for short period or distant systems where the expected astrometric wobble approaches or fall below a level of $\sim$\,0.1\,mas. \noise\ probably absorbs other source-related and artificial residue at present. The wobble of short period system more typical of the known LMXB population \citep[e.g. ][]{gandhi20} would be expected to be detectable only at small distances (Fig.\,\ref{fig:wobble}).

Accreting XRBs often comprise evolved donor stars undergoing Roche lobe overflow mass transfer that generates X-rays. Some CVs are similarly known to generate X-rays \citep[e.g. ][]{balman12}. We can thus look more closely at the HR diagram for sources off the MS and their known object classes from the literature. Fig.\,\ref{fig:distributions}c shows the \gaia\ colour--mag diagram with individual object classes denoted. Several classes immediately stand out, including binaries (B), YSOs (Y/T), variables (V), X-ray sources (V) and emission line objects (m). A handful of Wolf-Rayet systems (W) are also present at the luminous end. The largest `evolved' branch of primary sample sources includes 5\,$\pm$\,2\,\% of binaries, 21$_{-3}^{+4}$,\% of YSOs, 3$_{-1}^{+2}$,\% of variables and 5$_{-1}^{+2}$,\% of emission line objects, respectively. These fractions are higher than the corresponding values for the control sample. 

In total, the subsample lying outside of the MS comprises 283 (19\,$\pm$\,1\%) sources. Of these, 87 (31\,$\pm$\,4\%) have no published object classes, and are interesting sources for immediate follow-up. The mean $G$ mag of these sources is 15.2 (with a 1-$\sigma$ scatter of 1.3\,mag), so they are easily within reach of 2\,m class telescopes for spectroscopy. Radial velocity curves would be needed to test binarity and determine system characteristics, while deeper X-ray and radio data could establish the nature of the activity and help to confirm the presence and nature of any compact object components. Current model predictions suggest that \gaia\ ought to detect several thousand BHs in binary systems \citep{yamaguchi18, breivik17}, with a preference for more precise measurements of longer period systems \citep{andrews19}. Methods have also been proposed to detect non-interacting systems with MS companions \citep{shahaf19}. Quiescently accreting systems (of the kind that we have discussed herein) will likely be a fraction of these, but will be the `low hanging fruit' that are likely to be the easiest ones to follow up and confirm. In NS binaries, it may be possible to use high precision astrometry to determine compact object masses and constrain NS matter equation of state \citep{tomsick10}.

Similarly, sources classified only as emission line objects with little prior follow-up in this region of the HR plane could be hiding accreting systems with evolved donors. Variables, however, constitute a broad range of detailed source classes, and we cannot rule out the possibility that these may include single active stars where the variability causes systematic changes to the individual astrometric uncertainties resulting in artificially boosting \noise. 

The high abundance of YSOs could be a simple consequence of the fact that YSOs are thought to have a high multiplicity that decreases with evolutionary phase and scales with mass \citep[e.g. ][]{pomohaci19}. So younger `Y' systems in our primary sample could well include a high fraction of binaries. However, YSOs are known to be variable in flux, and are also known to be X-ray sources \citep{feigelson99}. YSOs may also be preferentially found in dense environments with additional ambient emission (e.g., clusters, molecular clouds, reflection nebulae). Though we have attempted to account for this by removing sources with near neighbours, we are unaware whether other artefacts might play a role in artificially boosting the numbers of these sources. Thus, both genuine binarity and systematic influences are also possible for YSOs. It should be stressed that the fractions of YSOs in the full sample remains small, much smaller than the numbers of known binaries and variables (cf. Table\,\ref{tab:fractions} and Fig.\,\ref{fig:typeshistall}); it is the relative fraction between primary vs. control that we are considering here. Similarly, we detect a few WR systems (a higher fraction in primary) that may also be subject to the effects of variability and binarity \citep[e.g. ][]{Vanbeveren98}.  

Our selection has ended up discovering a diverse range of source classes displaying a variety of `activity.' If astrometric noise selection is genuinely more effective in distinguishing these source classes, then the Galactic latitude distribution seen in Fig.\,\ref{fig:fullselectionsspatial} may be interpreted as a manifestation of their increasing fraction as one approaches the plane. Binaries and YSOs, for instance, correspond to younger evolutionary phases that would preferentially be found closer to their natal sites in the plane. Over time, these may diffuse outwards from the plane, resulting in the larger scatter in $z$ that we observe for the control sample. If true, this opens up a new astrometric pathway to probing distinct stellar populations over large swathes of the sky (cf. \citealt{belokurov20} for similar discussions using the RUWE parameter). We note that our sample is not meant to be complete in any physical sense yet. But understanding the astrometric selection function in detail will have to await future \gaia\ data releases.

\section{Conclusions}

We have explored the statistics and nature of objects found using astrometric excess noise selection in \gaia\ DR2. Our initial aim was to identify candidate  accreting systems. But a variety of tests carried out in a controlled fashion demonstrate that excess noise selection is effective in identifying a diverse range of active source classes. X-ray cross-matching is used to refine the selection to identify putative quiescent interacting binaries, variables, emission line sources and young stellar objects, amongst others. The sample also shows a high fraction of \ha-excess sources. 

Caution is needed when interpreting the measured values of excess noise, especially when \noise\ is small (well below 1\,mas). Full astrometric solutions in future data releases will help to understand the selection function more quantitatively. Upcoming all sky X-ray followup from \erosita\ will also provide a treasure trove of other candidate active systems enhancing the sample that we present here \citep{erosita}. Similarly, in the future, Galactic plane follow-up with the ngVLA should accomplish the same in the radio \citep{maccaronengvla}. 

We also release our full primary source sample to enable multiwavelength follow-up and characterisation of individual systems. 

\section*{Data availability}

All of the multiwavelength catalogues utilised herein (\gaia,  \chandra, and IPHAS) are publicly available. Our final extracted catalogue will also be available through CDS upon publication of this manuscript. Digital files for the SALT spectra will be available through the SALT archive, or from the authors upon request.

\section*{Acknowledgements}

PG acknowledges support STFC and a UGC-UKIERI Thematic Partnership. DAHB acknowledges research support from the South African National Research Foundation. PG is indebted to the {\em eROSITA} team, in particular J. Wilms and A. Schwope, for collaboration on a related expanded project. He also thanks J. Tomsick for discussion. 

This work has made use of data from the European Space
Agency (ESA) mission Gaia (https://www.cosmos.esa.
int/gaia), processed by the Gaia Data Processing and Analysis
Consortium (DPAC, https://www.cosmos.esa.int/
web/gaia/dpac/consortium). Funding for the DPAC has
been provided by national institutions, in particular the institutions
participating in the Gaia Multilateral Agreement. 

\chandra\ is a mission of the National Aeronautics and Space Agency (NASA). This research has made use of data obtained from the Chandra Source Catalog, provided by the Chandra X-ray Center (CXC) as part of the Chandra Data Archive.

Some of the observations presented here were obtained with SALT under programme 2018-2-LSP-001, which is supported by Poland under grant no. MNiSW DIR/WK/2016/07.

Extensive use was made of the {\tt TOPCAT} software \citep{topcat}, and the IDL {\tt astrolib} routines herein  \citep{idlastrolib}. This research has made use of the {\tt SIMBAD} database,
operated at CDS, Strasbourg, France, `The SIMBAD astronomical database' \citep{simbad}.




\bibliographystyle{mnras}
\bibliography{gandhi21} 



\clearpage
\setcounter{figure}{0}
\makeatletter 
\renewcommand{\thefigure}{A\@arabic\c@figure}
\makeatother
\setcounter{equation}{0}
\makeatletter 
\renewcommand{\theequation}{A\@arabic\c@equation}
\makeatother
\setcounter{table}{0}
\makeatletter 
\renewcommand{\thetable}{A\@arabic\c@table}
\makeatother
\setcounter{section}{0}
\makeatletter 
\renewcommand{\thesection}{A}
\setcounter{subsection}{0}
\makeatletter 
\renewcommand{\thesubsection}{A\@arabic\c@subsection}
\makeatother
\appendix

\section{Full catalogue}

A few example catalogue entries are listed in Table\,\ref{tab:fullcatalogue}. The full table will be available through CDS\footnote{\url{http://cdsportal.u-strasbg.fr/}}.

\begin{table*}
Complete X-ray matched primary sample catalogue\\
\begin{tabular}{lcccccccr}
\hline
\hline
RA$_{Gaia}$  &  Dec$_{Gaia}$  & $G$ & $d$ & \noise & $F_{\rm X}$ & $\Delta$ & Class & Notes\\
        deg  &  deg           & mag & kpc &   mas    &  10$^{-15}$\,erg\,s$^{-1}$\,cm$^{-2}$ & $''$ & &\\
\hline
39.059798816 &   59.692402778 & 15.19 & 1.700\,$\pm$\,0.134 & 0.25 & 4.4 & 0.40 & -- & --\\
53.243898542 &  -27.835516974 & 17.44  &  0.448\,$\pm$\,0.031 & 1.01 & 0.11 & 0.26 & -- & Incorrectly classed as Galaxy in {\tt SIMBAD}\\
80.205671349 &  -45.691612742 & 11.47 & 0.103\,$\pm$\,0.002 & 1.04 & 1063 & 0.43 & BY\,Dra & --\\
\hline
\end{tabular}
\caption{ICRS \gaia\ source coordinates are listed.  $d$ here is computed from simple parallax inversion.   $d$ here is computed from simple parallax inversion.  $\Delta$ is the separation between the \gaia\ and \chandra\ counterparts. A portion of the catalogue is shown here for reference. $F_{\rm X}$ denotes the CSC broadband flux (0.5--7\,keV).\label{tab:fullcatalogue}}
\end{table*}


\section{A simulation based on DR2}

\begin{figure*}
\hspace*{-1cm}
	\includegraphics[angle=0,width=1.\columnwidth]{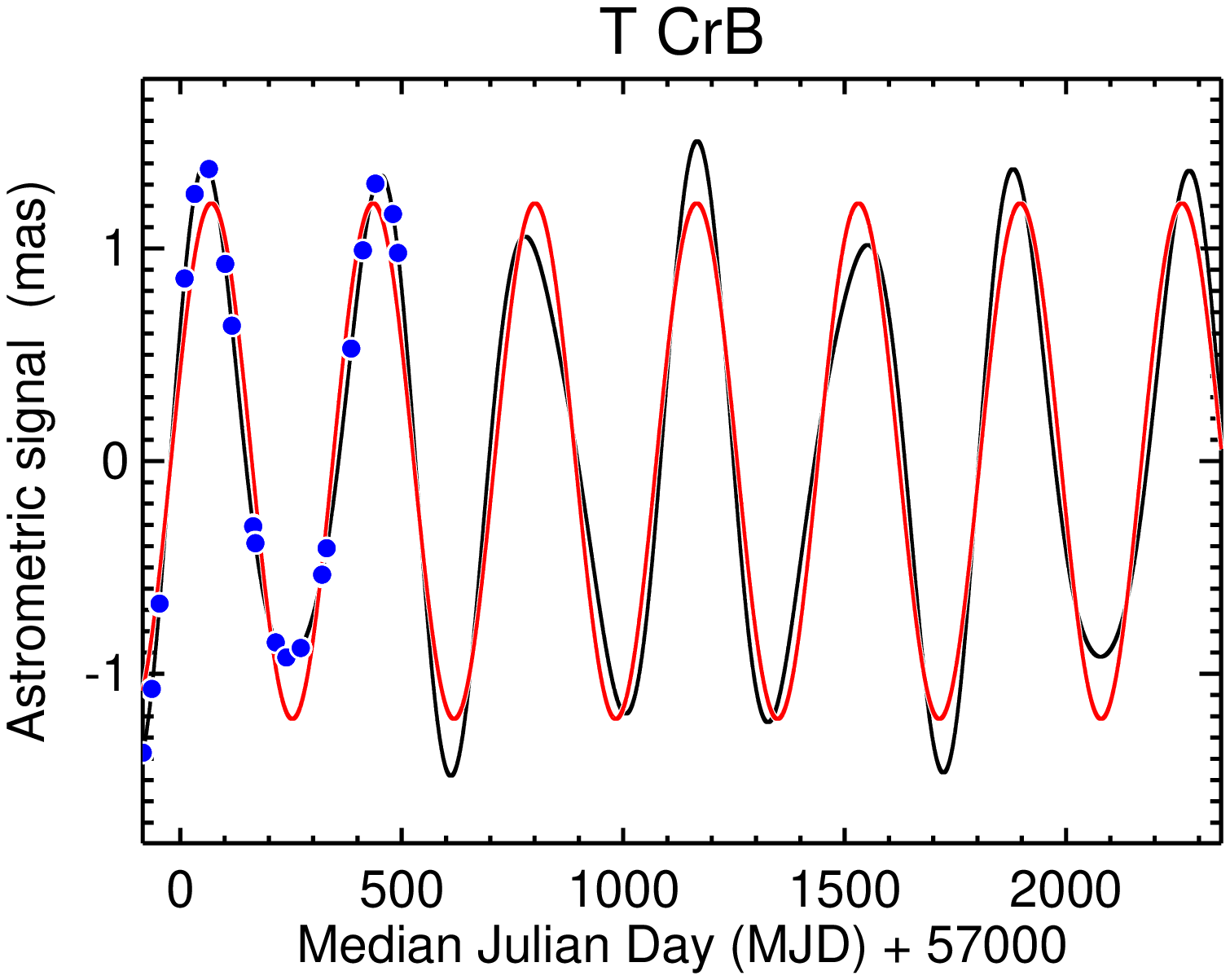}
    	\includegraphics[angle=0,width=1.\columnwidth]{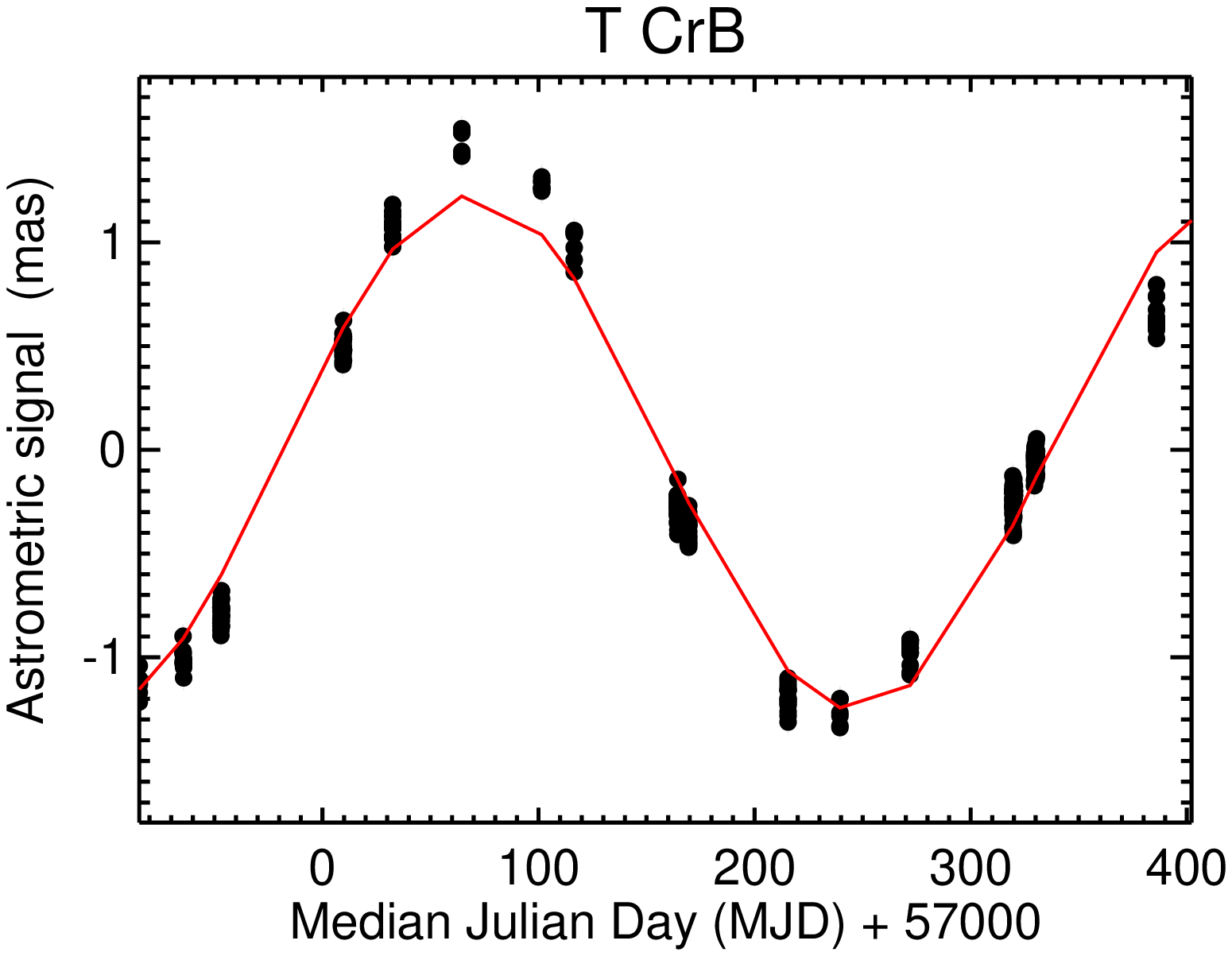}
	\includegraphics[angle=0,width=1.\columnwidth]{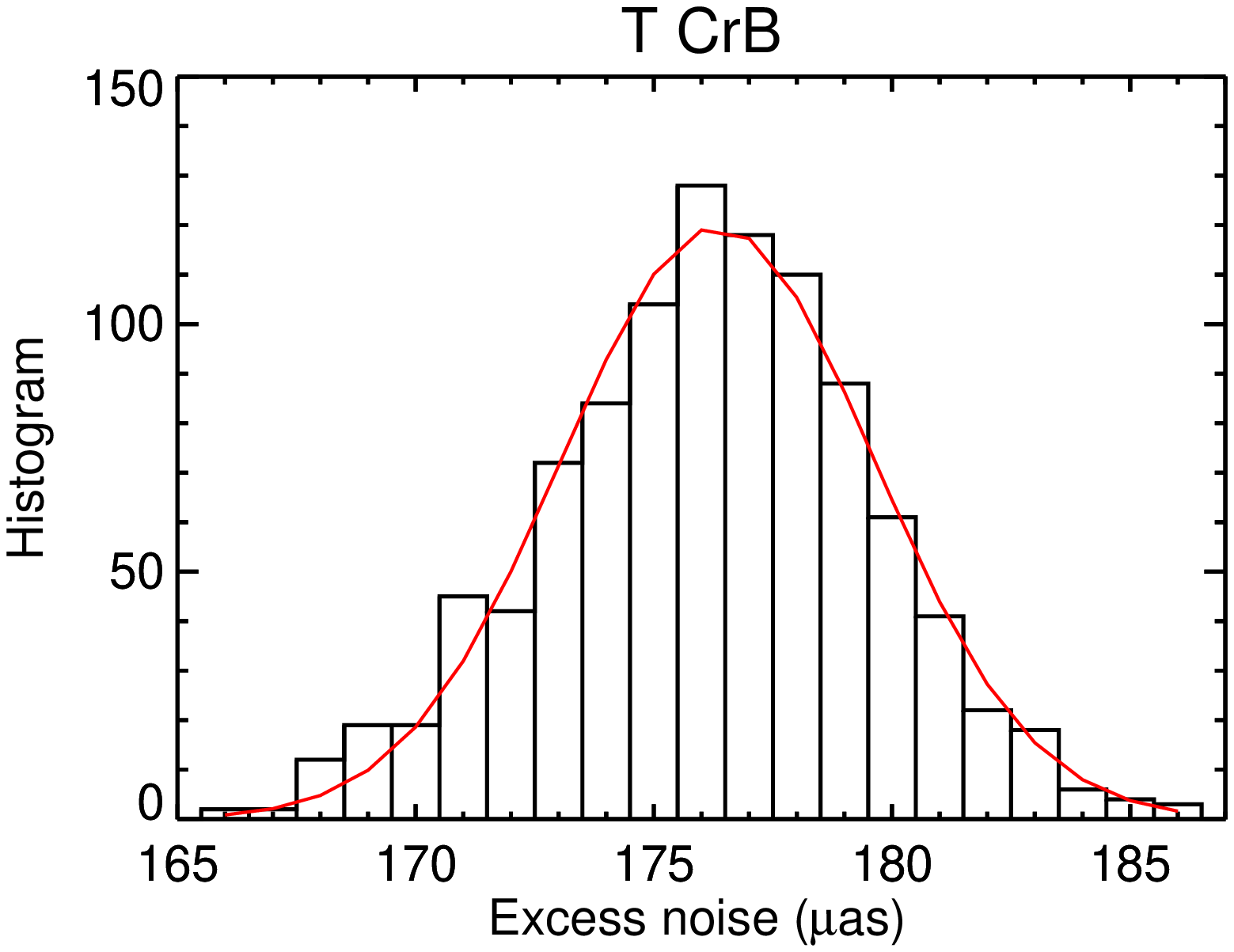}
\caption{
Schematic illustration of parallax (red) and orbital wobble (black) for the case of T\,CrB, over a $\approx$\,6-year period. The blue points denote the 51 forecasted observation times of the source using the \gaia\ Observation Scheduling Tool (GOST) over the 22\,month period of \gaia\ DR2 from mid-2014 to mid-2016. 
(b) Right: Simulation of observations of T\,CrB based upon the dates returned by the \gaia\ Observation Schedule Tool (GOST). There are 51 field-of-view transits over the DR2 period, with each transit taken to comprise 8 CCD observations. Our simulations assume Normal distribution sampling together with  magnitude dependent uncertainties (see \S\,\ref{sec:understanding}). (c) Bottom: Astrometric excess noise (\noise) distribution based upon 1,000 simulations. For comparison, the reported DR2 value of \noise\ for T\,CrB is 180\,$\pm$\,10\, $\mu$as. 
    \label{fig:v404sim}}
\end{figure*}

In order to gain more insight into the meaning of orbital wobble and astrometric noise, we carried out simulations of \gaia\ observations of one of these XRBs, T\,CrB. This is illustrated in the example schematic in Fig.\,\ref{fig:v404sim}a, which shows the orbital wobble \w\ signature superposed on the annual parallax ($\pi$) swing. 
This source possesses a well known ephermeris with \porb\,=\,227.57\,days and with a binary separation $a$\,=\,0.54\,AU \citep{fekel00}. The DR2 parallax of $\pi$\,=\,1.21\,mas corresponds to a distance of 0.8\,kpc \citep{bailerjones2018}, yielding a maximum angular size ($\theta_a$) of the binary orbit of about 0.68\,mas. Caution is warranted when deriving physical parameters for objects such as T\,CrB where $\theta_a$ is a significant fraction of $\pi$ (cf. also \citealt{schaefer18}). However, our purpose here is to illustrate the expected magnitude of \noise\ through approximate simulations; we do not intend to infer accurate system parameters. 

The astrometric wobble $\omega$ is expected to be smaller than $\theta_a$ because both components of the binary are luminous in this case, so the effective photocentre lies in between the two components. $\omega$ can be computed with knowledge of the mass ratio ($q$) and luminosity ratio ($l$) of the binary components \citep{belokurov20}, both of which are known for T\,CrB: $q$\,=\,0.82 and $l$\,=\,6.55 \citep{stanishev04, schaefer09}, though this $l$ reflects the bolometric luminosity ratio, not the ratio as seen in the \gaia\ $G$ band, and note that irradiation of the donor by the accretion disc is also likely to be important \citep[e.g. ][]{hachisu99}. 

This yields a wobble $\omega$\,=\,0.28\,mas, and an expectation value $\hat{\omega}$\,=\,0.20\,mas. Encouragingly, \gaia\ DR2 reports a very similar excess noise term \noise\,=\,0.18\,$\pm$\,0.01\,mas. 

The recovery of this value can be verified using simulations of astrometric fitting. The standard astrometric pipeline fits position, proper motion and parallax to the observation set. There were 51 
forecasted transits across the \gaia\ field-of-view for T\,CrB during the 22 month period of DR (2014 July -- 2016 May).\footnote{\url{https://gaia.esac.esa.int/gost/}}
 Each transit comprises multiple CCD observations as the source drifts across the detector field of view. We assumed 8 observations per transit, as there are 9 consecutive CCDs in the focal plane \citep{gaiamission}, but not every CCD is covered in each transit. This results in 408 
 assumed observations, as compared to the true number of observations ({\tt n\_observations}\,=\,414 
 reported in DR2)\footnote{The number of good observations is marginally lower, {\tt astrometric\_n\_good\_obs\_al}\,=\,396}, which suffices for our purposes of an illustrative simulation. 
 The proper comparison will only be possible once the full observation and solution set is released.\footnote{\url{https://www.cosmos.esa.int/web/gaia/release}} 

Each observation was simulated using the ephemeris and the forecasted dates. 
The formal uncertainties were derived from the reported precision of the along-scan astrometric measurements as a function of source magnitude by \citet[][]{lindegren18};  see also Fig.\,\ref{fig:wobble}. Simulated astrometric signals were drawn from a Normal distribution centred on the expected orbital phase at the forecasted date, with a scatter given by the above uncertainties. These were then fitted with a 4-parameter function including offsets representing position, proper motion and phase offsets as well as the parallax amplitude. Our simulations are taken to be one-dimensional, focusing on the overall excess astrometric signal for simplicity, instead of decomposing this into its two-dimensional components related to the source position in the plane of the sky. This is, of course, an approximation of the full 5-parameter multivariate fit carried out by the DR2 pipeline, but nevertheless is useful for understanding the limits of the data. 

An example astrometric fit to one of the simulations is shown in Fig.\,\ref{fig:v404sim}b. \noise\ was determined as the excess noise floor required to make the reduced $\chi^2$ statistic equal 1 for a suite of 1,000 simulations, and the resultant distribution of \noise\ is shown in Fig.\,\ref{fig:v404sim}c. We find a mean \noise\,=\,0.17\,mas, not far off the expectation value and the DR2 value stated above. 

This match verifies that the expected pipeline flow and the number of observations can approximately reproduce the wobble signature in this case. It is meant to be illustrative and is not surprising. More surprising are the mismatches noted in \S\,\ref{sec:noisemag}, where \noise\ far exceeds $\hat{\omega}$. These would require additional intrinsic or artefact noise terms to explain.

\section{SIMBAD Object Classifications and Assigned Codes}

Tables\,C1 and C2 list the individual source types (the {\tt main\_type} from {\tt SIMBAD}) together with the corresponding assigned short code denoting the source category used in the Results, Discussion and some of the figures presented herein.

We caution that a few extragalactic source types are also listed. We have cross-checked a few of the relevant sources individually and believe their {\tt SIMBAD} types to be erroneous, given the significant parallax detections. These constitute a very small fraction of the overall samples; nevertheless, these require further confirmation. 

{\centering
\begin{table}
\centering
{\small
Object classifications from SIMBAD\\
\begin{tabular}{r|r}
\hline
\hline
Class  & Assigned Code\\
\hline
                           &\\
                    Unknown&\\
           multiple\_object&\\
                        **&*\\
                     *in**&*\\
                    *inNeb&*\\
                      AGB*&*\\
                   BlueSG*&*\\
                        C*&*\\
           Candidate\_AGB*&*\\
            Candidate\_HB*&*\\
            Candidate\_Hsd&*\\
           Candidate\_RGB*&*\\
           Candidate\_RSG*&*\\
             Candidate\_S*&*\\
        Candidate\_brownD*&*\\
      Candidate\_low-mass*&*\\
      Candidate\_post-AGB*&*\\
                       HB*&*\\
                       PM*&*\\
                      Pec*&*\\
                      RGB*&*\\
                    RedSG*&*\\
                        S*&*\\
                       SG*&*\\
                      Star&*\\
                   brownD*&*\\
                 low-mass*&*\\
                 post-AGB*&*\\
                       AGN&o\\
            AGN\_Candidate&o\\
                     BYDra&B\\
            Candidate\_CV*&B\\
            Candidate\_EB*&B\\
            Candidate\_XB*&B\\
                 CataclyV*&B\\
                     DQHer&B\\
                 DwarfNova&B\\
                       EB*&B\\
                  EB*Algol&B\\
                   EB*WUMa&B\\
                 EB*betLyr&B\\
                      HMXB&B\\
                      LMXB&B\\
                      Nova&B\\
                 Nova-like&B\\
                     RSCVn&B\\
              RotV*alf2CVn&B\\
                       SB*&B\\
                Symbiotic*&B\\
                   BClG&BClG\\
                   Blue&Blue\\
 BlueStraggler&BlueStraggler\\
             Candidate\_C*&C\\
                     Cloud&o\\
            Compact\_Gr\_G&o\\
                     DkNeb&o\\
                       EmG&o\\
                       HII&o\\
                 IR$<$10um&I\\
                 IR$>$30um&I\\
                       IR&IR\\
                    MolCld&o\\
                     OH/IR&o\\
\hline
\end{tabular}
\caption{\label{tab:classes}}
}
\end{table}
\begin{table}
\centering
{\small
Object classifications from SIMBAD (contd.)\\
\begin{tabular}{r|r}
\hline
\hline
Class  & Assigned Code\\
\hline
                      OpCl&o\\
                     PairG&o\\
               PartofCloud&o\\
                       QSO&o\\
                 Radio(cm)&o\\
                 Radio(mm)&o\\
             Radio(sub-mm)&o\\
                       Red&o\\
                     RfNeb&o\\
                 Radio&Radio\\
          Candidate\_TTau*&T\\
                     TTau*&T\\
                       UV&UV\\
            Candidate\_LP*&V\\
            Candidate\_Mi*&V\\
          Candidate\_RRLyr&V\\
                   Cepheid&V\\
                  EllipVar&V\\
                Erupt*RCrB&V\\
                 Eruptive*&V\\
                    Flare*&V\\
                       HV*&V\\
             Irregular\_V*&V\\
                      LPV*&V\\
                      Mira&V\\
                 Orion\_V*&V\\
                    PulsV*&V\\
               PulsV*RVTau&V\\
                PulsV*WVir&V\\
                PulsV*bCep&V\\
              PulsV*delSct&V\\
                     RRLyr&V\\
                     RotV*&V\\
                 Transient&V\\
                        V*&V\\
                       V*?&V\\
                  deltaCep&V\\
                  gammaDor&V\\
                  pulsV*SX&V\\
           Candidate\_WD*&WD\\
           Candidate\_WR*&W\\
                      WD*&WD\\
                      WR*&W\\
                         X&X\\
            Candidate\_YSO&Y\\
                       YSO&Y\\
                     gamma&g\\
                gammaBurst&g\\
                       Ae*&m\\
                       Be*&m\\
            Candidate\_Ae*&m\\
            Candidate\_Be*&m\\
                       Em*&m\\
                     EmObj&m\\
                     BLLac&o\\
          BLLac\_Candidate&o\\
                    Galaxy&o\\
                     GinCl&o\\
                  GinGroup&o\\
                        PN&o\\
                       PN?&o\\
                    Planet&o\\
                   Planet?&o\\
           Candidate\_pMS*&p\\
                      pMS*&p\\
              HotSubdwarf&o\\
\hline
\end{tabular}
\caption{\label{tab:classescontd}}
}
\end{table}
}

\section{Selection on RUWE}

\begin{figure}
    \centering
    \includegraphics[width=8.5cm]{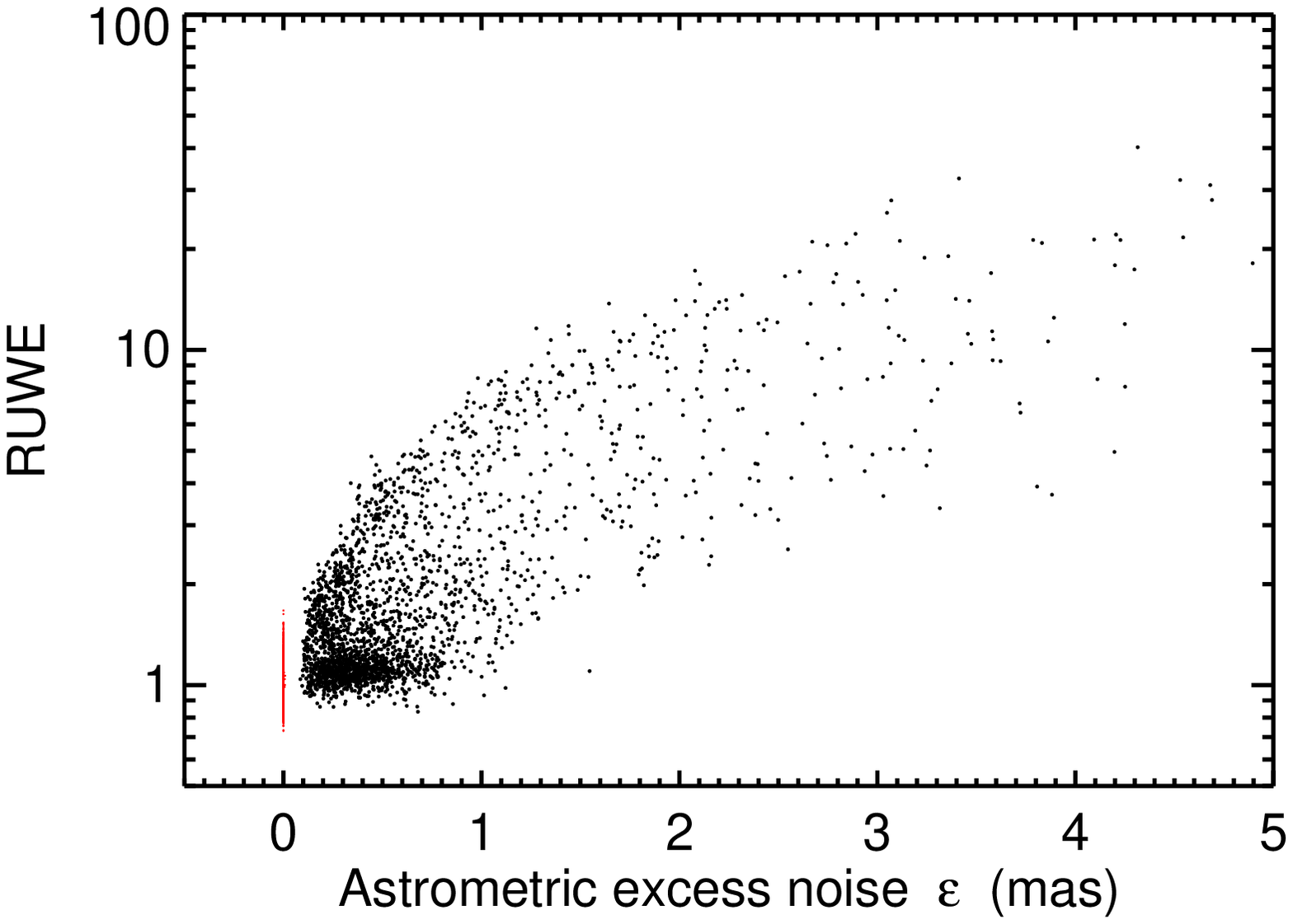}
    \caption{\noise\  (astrometric excess noise) vs. RUWE for the primary (black) and control (red) samples.}
    \label{fig:wruwe}
\end{figure}

After DR2 release, identification of some issues with the data pipeline led to the definition of a new fit parameter statistic termed RUWE (Renormalised Unit Weight Error). This attempts to correct for issues such as a degrees-of-freedom bug as well as systematic fit variations based upon colour \citep{lindegren18}. A good astrometric single-source solution has an expectation value $\widehat{\rm RUWE}$\,=\,1. Significantly higher values are indicative of other factors, including artefacts or intrinsic source complexity. Thus, RUWE is expected to scale with \noise. Fig.\,\ref{fig:wruwe} confirms this, with our primary sample showing a substantial tail to higher RUWE values: $\langle {\rm RUWE}\rangle$\,=\,2.74 and  $\langle {\rm RUWE}'\rangle$\,=\,0.99, respectively. RUWE selection and \noise\ selection are thus complementary to each other, with each having its pros and cons \citep[e.g., ][]{belokurov20,penoyre20}. Here, we have focused on \noise\ due to its straightforward statistical interpretation which could also, in principle, be translated into physically interesting system properties. Moreover, the DR2 pipeline quantifies the significance of \noise, unlike RUWE. Finally, the aforementioned pipeline bugs (see Section\,\ref{sec:fullsample}), at most, result in an {\em underestimate} of \noise, so our selection is likely somewhat conservative.

\bsp	
\label{lastpage}
\end{document}